\documentclass[pre,twocolumn,superscriptaddress,floatfix,nofootinbib]{revtex4-1}  

\usepackage[paperwidth=210mm,paperheight=297mm,centering,hmargin=2cm,vmargin=2.5cm]{geometry}

\usepackage[pdftex]{graphicx}
\usepackage{amsmath,amsfonts,amssymb}
\usepackage{natbib}
\usepackage{MnSymbol}
\usepackage{hyperref}
\usepackage{xcolor} 
\usepackage{array}
\usepackage{pgfplots}
\usepackage{graphicx}
\usepackage[left,modulo,pagewise]{lineno}
\usepackage{multirow}
\usepackage{csquotes}

\definecolor{bluemoi}{rgb}{0.25,0.50 ,0.75} 

\hypersetup{
  bookmarksopen=true,
  pdftitle=" ",
  pdfauthor=" ", 
  pdfsubject=" ", 
  pdftoolbar=true, 
  pdfmenubar=true, 
  pdfhighlight=/O, 
  colorlinks=true, 
  pdfpagemode=UseNone, 
  pdfpagelayout=SinglePage, 
  pdffitwindow=true, 
  linkcolor=bluemoi, 
  citecolor=bluemoi, 
  urlcolor=bluemoi 
}

\makeatletter
\renewcommand{\figurename}{\sf \textbf{Figure}}
\renewcommand{\thefigure}{\arabic{figure}}
\renewcommand{\fnum@figure}{\sf\textbf{\figurename}~\textbf{\thefigure}}
\renewcommand{\tablename}{\sf\textbf{Table}}
\renewcommand{\thetable}{\arabic{table}}
\renewcommand{\fnum@table}{\sf\textbf{\tablename}~\textbf{\thetable}}
\makeatother

\begin{document}

\title{Immigrant community integration in world cities}

\author{Fabio Lamanna}
\affiliation{Instituto de F\'isica Interdisciplinar y Sistemas Complejos IFISC (CSIC-UIB), Campus UIB, ES-07122 Palma de Mallorca, Spain}

\author{Maxime Lenormand}
\affiliation{Irstea, UMR TETIS, 500 rue JF Breton, FR-34093 Montpellier, France}

\author{Mar\'ia Henar Salas-Olmedo}
\affiliation{Departamento de Geograf\'ia Humana, Facultad de Geograf\'ia e Historia, Universidad Complutense de Madrid, 28040, Madrid, Spain}

\author{Gustavo Romanillos}
\affiliation{Departamento de Geograf\'ia Humana, Facultad de Geograf\'ia e Historia, Universidad Complutense de Madrid, 28040, Madrid, Spain}

\author{Bruno Gon\c calves}
\affiliation{Center for Data Science, New York University, New York, 10011 NY, USA}

\author{Jos\'e J. Ramasco}
\thanks{Corresponding author: jramasco@ifisc.uib-csic.es}
\affiliation{Instituto de F\'isica Interdisciplinar y Sistemas Complejos IFISC (CSIC-UIB), Campus UIB, ES-07122 Palma de Mallorca, Spain}

\begin{abstract} 
As a consequence of the accelerated globalization process, today major cities all over the world are characterized by an increasing multiculturalism. The integration of immigrant communities may be affected by social polarization and spatial segregation. How are these dynamics evolving over time? To what extent the different policies launched to tackle these problems are working? These are critical questions traditionally addressed by studies based on surveys and census data. Such sources are safe to avoid spurious biases, but the data collection becomes an intensive and rather expensive work. Here, we conduct a comprehensive study on immigrant integration in $53$ world cities by introducing an innovative approach: an analysis of the spatio-temporal communication patterns of immigrant and local communities based on language detection in Twitter and on novel metrics of spatial integration. We quantify the {\it Power of Integration} of cities --their capacity to spatially integrate diverse cultures-- and characterize the relations between different cultures when acting as hosts or immigrants.
\end{abstract}

\maketitle

\section*{Introduction}

Immigrant integration is a complex process involving a multitude of aspects such as religion, language, education, employment, accommodation, legal recognition and many others. Its study counts with a long tradition in sociology through concepts such as immigrant assimilation \cite{Burgess1921}, structural assimilation \cite{Gordon1964} or immigrant acculturation and adaptation \cite{Berry1997}. Over the last years, there have been advances in the definition of a common framework concerning immigration studies and policies \cite{Ager2008}, although the approach to this issue remains strongly country-based \cite{Entzinger2005}. The outcome of the process actually depends on the culture of origin, the one of integration and the policies of the hosting country government \cite{Gonul}. Traditionally, spatial segregation in the residential patterns of a certain community has been taken as an indication of ghettoization or lack of integration \cite{Massey1993}. While this applies to immigrant communities, it can also affect to minorities within a single country \cite{Massey1987}. The spatial isolation reflects in the economic status of the segregated community and in social relationships of its members \cite{Oka2015}.

In global terms while international migration flows have remained almost stable over the last 20 years \cite{Abel2014,Butler2017}, political and economic upheavals such as the Arab Spring and the Syrian civil war have brought the problem of migrants and their integration to the forefront of world news and even the academic press \cite{editorial2017,Dijstelbloem2017}. A good part of newcomers concentrates in cities, and particularly in the large metropolises known as World Cities. These are centers that attract specialized immigration, driving important social and cultural transformations in cities worldwide \cite{Beaverstock1996}. The concept of Global or World Cities emerged in the 80s \cite{Sassen2005, Friedmann1986} as strategic territories that articulate the international economic structure. According to Sassen \cite{Sassen2005}, Global Cities are not only characterized by growing multiculturalism but also by a rising social polarization, which was finally materialized into an increasing social spatial segregation and gentrification processes. This assertion is still under debate in the area of social sciences, requiring 
its settlement further empirical evidence \cite{Samers2002, Hamnett1994}. Furthermore, immigrant integration has been the focus of many research studies, most of which conducted from national perspectives especially in European countries and the USA \cite{Entzinger2005,Gonul,Massey1987,Musterd2005,Bean2003}, and it is still in dare need of information sources beyond national census \cite{editorial2017,Dijstelbloem2017,Phalet2003}.

In parallel, the last few years have brought a paradigm shift in the context of socio-technical data. Human interactions are being digitally traced, recorded and analyzed in large scale. Sources as varied as mobile phone records \cite{Reades2007,Gonzalez2008,Reades2009,Soto2011,Toole2012,Pei2013,Louail2014,Amini2014,Tizzoni2014,Deville2014,Grauwin2014,Blondel2015,Louail2015}, credit card transactions \cite{Lenormand2015c}, or Twitter data \cite{Hawelka2014,Lenormand2014,Lenormand2014b} have been used to study mobility and land use in urban areas. Most of these works have been carried out in the zones where data was available, mostly inside cities or single countries. Twitter data has, however, the particularity of extending beyond national borders and, therefore, it allows researchers to analyze mobility and city hierarchies at an international level \cite{Hawelka2014,Lenormand2015}. Besides activity and mobility, the content of the tweets bears also a wealth of information starting by the language in which the text is written. 
The spatial distribution of languages has been investigated in Refs. \cite{Magdy2007,Mocanu2013,Jurdak2014}, exploring as well the relations between languages trough multilingual individuals, and in Refs. \cite{Goncalves2014,Doyle2014}, where the spatial extension of Spanish and English dialects was examined. Of course, one of the weak points of Twitter as data source is its representativeness. This question has been boarded in Refs. \cite{Mislove2011,Lenormand2014,Bokanyi2016,Sloan2017}, finding acceptable coverage for the American, British and Spanish populations in terms of geographic allocation, race, religion and mobility, although the data shows a bias towards younger individuals. In this context, it is of special interest the mix of location and language detection. This issue opens the door to characterize foreign users in short visits, temporal or permanent stays. Arribas-Bel \cite{Arribas-Bel2015} published a first exploratory work on this direction using Twitter and census data in Amsterdam. Contemporarily, the use of phone call records to foreign countries has provided a picture of communities with external connections in the area of Milan \cite{Bajardi2015}. When it comes to immigrant integration, there are less works but one that deserves mention is a study recently published by \cite{Mason2016} who looked at the social ties (friendships and affinities) between immigrant communities by using data from Facebook. There have been diverse attempts to measure the degree of immigrant integration over the last years \cite{Vigdor2008} by introducing a quantitative index, the Composite Assimilation Index (CAI), that quantifies the degree of similarity between native- and foreign-born adults in the United States, based on US census data. In \cite{Mason2016}, a similar measure of integration is considered based on the relative proportion of ties between immigrant people born in the US, compatriots living in the US, and inter-group friendships with immigrants from other countries.

In this work, we introduce a novel approach to quantify the spatial integration of immigrant communities in urban areas worldwide. By analyzing language in Twitter data, we are able to assign languages to each user paying special attention to those corresponding to migrant communities in the city considered. The individuals' digital spatio-temporal communication patterns allow us to define as well areas of residence. With this information, we perform a spatial distribution analysis through a modified entropy metric, as a quantitative way to measure the spatial integration of each community. The metric can be expressed in a bipartite network with the culture of origin in one side and the hosting cities, countries and languages in the other. These results lead us to categorize the cities according to how well they integrate immigrant communities and also to quantify how well hosting countries integrate people from other cultures. 

\begin{figure*}
	\includegraphics[width=\linewidth]{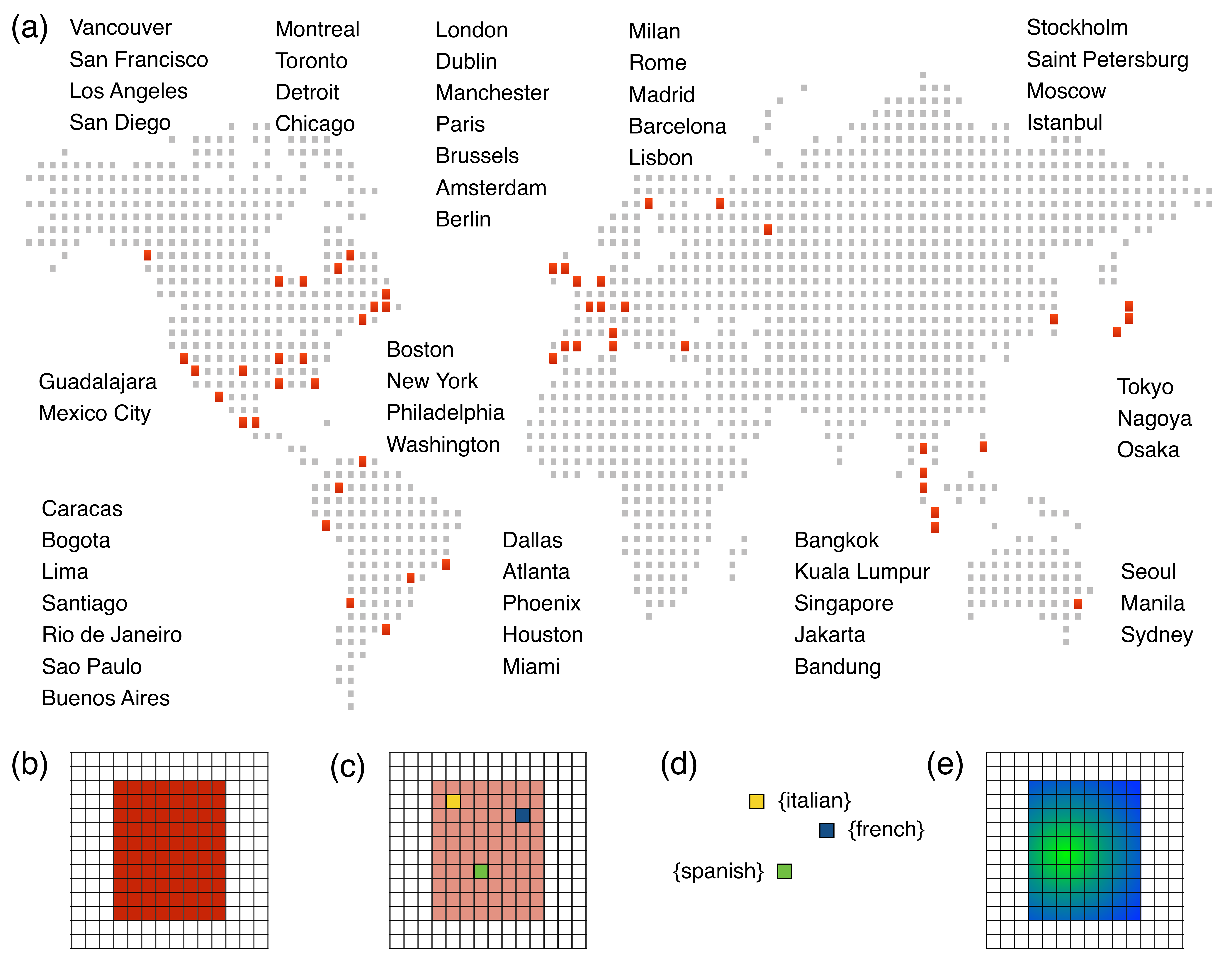}
	\caption{\textbf{Dataset and framework description}. The cities passed through the lens of our analysis are mostly distributed over four continents (a). Africa has been not considered due to the lack of data. We cover each city with a square grid in order to keep a homogeneous spatial division over the whole urban area where the users are going to be distributed (b), selecting resident users and their most frequent location thanks to their activity over space and time (c). In addition, we assign the users' most probable native language (d) and perform a spatial analysis over the cities (e) to get information about the population distribution in function of the language spoken by the users.}
	\label{fig:fig1}
\end{figure*}

\section*{Materials and methods}

We selected $53$ of the most populated cities in the world (see Figure \ref{fig:fig1}a) and analyzed the geo-localized tweets originating in each city between October 2010 and December 2015 as captured from the Twitter API. The data was collected respecting Twitter's terms of service and privacy conditions. Several items are extracted from each tweet: user ID, geographical coordinates (latitude and longitude), date and time and the text of the tweet. In order to get a coherent picture in the different time zones, we convert the Twitter UTC time into the local timezone for each city. Before starting with the analysis, it is necessary to filter out non-human users from the  dataset. This is fundamental in order to prevent result pollution by signals coming from automatic tweet generators (bots), which are not rare in social networks \cite{Chu2010}. We found and disregarded tweets generated at the same time (with the precision of the second) by the same account. Moreover, we discard users who tweet more than three times per minute. Finally, we detect the speed of users moving through consecutive locations in order to filter out those traveling faster than a reasonable speed in urban areas (100 km/h or 62 mph). This procedure leaves us with a total of $350.9$ millions of tweets posted by $14.5$ millions of users in the $53$ cities (see Table S1 in Appendix for detailed numbers per city).

We will propose below a metric to assess spatial segregation of immigrant communities that is not highly sensitive to the specific borders of the area studied. However, everything has its limits. The mix of local and immigrant population is different in urban and rural areas. It is important thus to attain a balance and ensure that the region considered contains the city, where the signal on immigrants is stronger, but it does not extend unnecessarily far from it. This means that we should agree on a city definition that can be applied around the world and it is large enough to include the whole metropolitan area. Unfortunately, generic definitions such as the Larger Urban Zone (LUZ) definition of Eurostat for Europe does not exist at the global scale. There are plenty of different ways of defining cities, with, for example, methods based on urban growth, percolation, attraction or fractal theory. All these methods require third party data such as population, built-up area or flows of commuters that is not easily available in a consistent form everywhere. To side step this difficulty, we use a very pragmatic definition based only on the Euclidean distance and consider all activity within a frame of $60 \times 60$ km$^2$ centered on the barycenters listed in Table S2 in Appendix to belong to the city itself, dividing each city area using an equally spaced grid of $500 \times 500$ meters (Figure \ref{fig:fig1}b).

\begin{figure}
\centering 
\includegraphics[width=\linewidth]{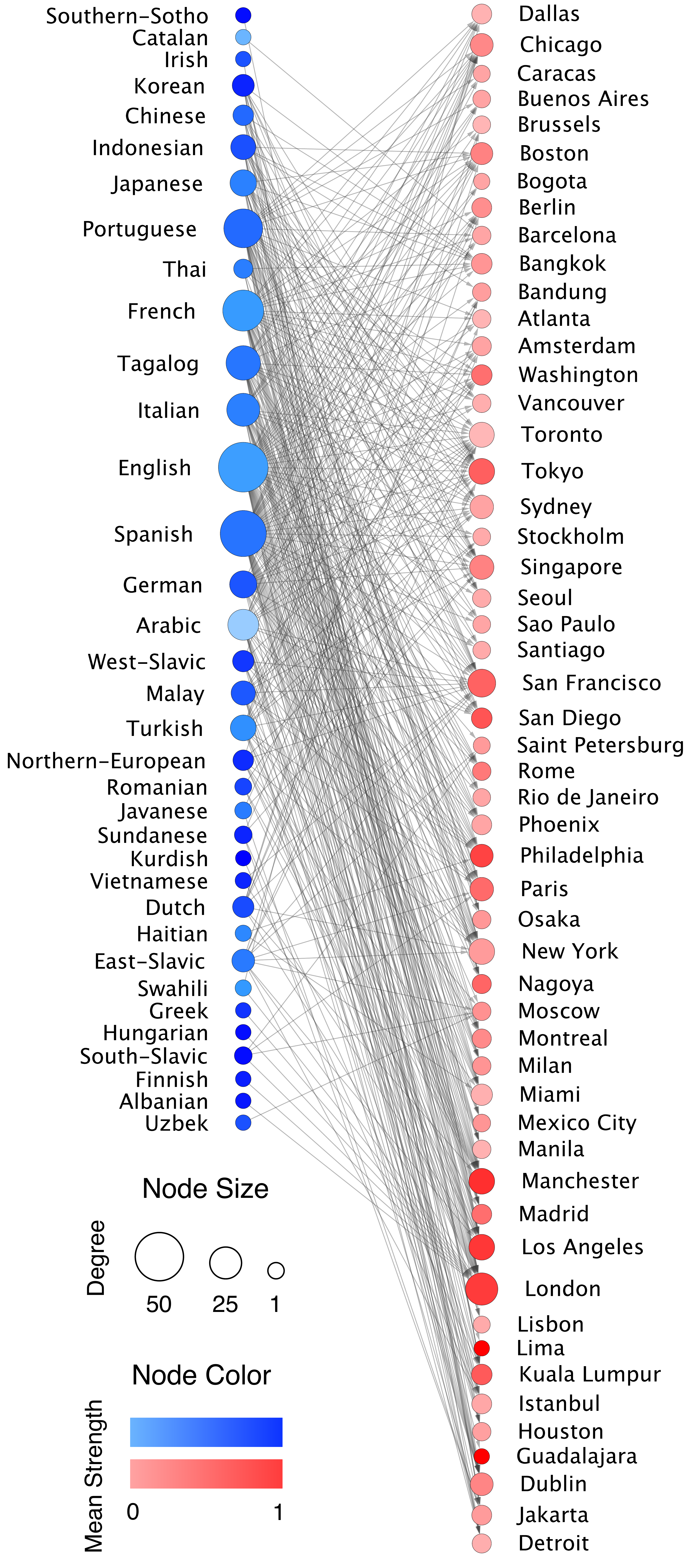}
\caption{\textbf{Bipartite Spatial Integration Network}. The network comprises of two sets: \textbf{L} of Languages and \textbf{C} of cities; the languages detected are connected to the cities set where the corresponding community of immigrants has been found. The weight of the edge corresponds to the values of $h_{l,c}$. The size of the nodes is proportional to its degree and the color to its mean strength.}
\label{fig:fig2}
\end{figure}

\subsection*{Definition of the user's place of residence}

As represented in Figure \ref{fig:fig1}c, the place of residence of every user is defined as the most frequented grid cell between 8pm and 8am local time. To ensure that a user shows enough regularity and that he/she is really living in the city, and not just a visitor for a small period of time, we applied three filters: a minimum number of consecutive months of activity $C$, a minimum number of hours spent by the user in the most frequented cell $N$ measured out of his/her consecutive tweets, and $\Delta$ as the ratio between $N$ and the total number of hours of activity for each user (number of hours during which he/she has posted at least one tweet). The source code used to extract most visited locations from individual spatio-temporal trajectories is available online\footnote{\url{https://github.com/maximelenormand/Most-frequented-locations}}.

Users who are active within a given city for at least three consecutive months are considered to be residents, so this establishes the first condition $C \ge 3$ months. The values of the other two parameters were determined empirically. In Figure S1 in Appendix, we plot the evolution of the number of users left in the dataset as a function of $\Delta$ for different values of  $N = [5,10,15,20]$  in each of the 53 cities. As the shape of the curves is similar for different values of $N$, it does not seem to be a natural features that would allow us to define a clear cutoff. We fix $\Delta \ge 0.2$ and $N \ge 5$, as a trade-off between being relatively sure about the users' residence area and keeping enough number of users to have proper statistics. Table S3 in Appendix lists the final number of residents per city after this data cleaning procedure. Note that there are at least $1000$ reliable users per city.

\subsection*{Language assignment} 

At this point, we are interested in introducing a method to determine which languages each user speaks, or at least in which languages he/she tweets. If any of these languages is proper of an immigrant community, this most likely will identify the user as a member of that community. To do this, the language in each tweet is detected using the version 2.0 of the \enquote{Chromium Compact Language Detector} (CLD2), which returns the languages detected along with a confidence assessment. CLD2 implements a Bayesian classifier for detecting language from UTF-8 text. Twitter entities (urls, mentions, hashtags) that may difficult our language detection efforts are removed, and only the remaining text was given as input to CLD2. To obtain reliable results, we keep only tweets for which the detector returned a language with confidence level of at least $90 \%$. Also, we aggregate close languages to take into account the uncertainty in the identification of \enquote{mutually intelligible} languages and dialectal varieties (see Table S4 in Appendix for more details). 

As can be expected, there are users tweeting in more than one language. We create a dictionary of the occurrences of each language in each users' tweets pattern. English is one of the most frequent language per user, because of its diffusion as \enquote{lingua franca} for spreading information to the highest number of Twitter followers. Still since we are interested in finding the language representative of users' community of origin, we propose a language algebra in order to extract this information from the user's dictionary. Let us define as \(Local\) the official language of each city. There are cases where there can be more than one Local language coexisting in the same city, like Catalan and Spanish in Barcelona, French and Flemish in Brussels or French and English in Montreal. The same occurs for Dublin and Singapore (see Table S5 for a complete list of cities and languages). After defining the Local languages in each city, we assign to each user its most frequent language. In case of bilingual/multilingual users, we set as user's language the one which differs from English or the Local unless these are the only two languages in the dictionary. In this latter case, we define the user as speaker of the Local language. In case of three languages spoken by the same user, we adopted the same hypothesis, assigning to the user the \enquote{third} language spoken apart when only one or both between English and Local are in the dictionary. In general, we take the most popular language in the dictionary other than English and the Local ones. If there are only Local languages and English, we keep the Local. English can be only assigned if it is the only one in the dictionary. The final number of users left for the analysis with a reliable residence cell, per language identified and per city are displayed in Table S6 in Appendix. We consider languages in each city with {\bf $30$} users or more.

\section*{Results}

\subsection*{Bipartite spatial integration network}

To quantify the spatial segregation of each immigrant community in every city, we build a bipartite spatial integration network $H$ (see Figure \ref{fig:fig2}). Every language is connected to the cities where the corresponding immigrant communities has been detected. The weight of an edge between language $l$ and city $c$, $h_{l,c}$, corresponds to the level of spatial integration measured with a new metric inspired by the Shannon entropy, but modified to take into account the finite character of the sampling of communities in our Twitter database. Shannon entropy-like descriptors have been used before in this context especially when considering the spatial segregation of ethnic minorities in the US cities \cite{White1986}. Recalling that the cities have been divided in equal area grid cells and focusing first only on one generic city $c$, we can directly calculate from the data the fraction of users of a certain community $l$ having their residence at cell $i$, $p_{l,i}$. This allows us to define an entropy per language community $l$: 
\begin{equation}
\label{entropy}
s_{l,c} = - \sum_{i=1}^{N} p_{l,i} \, \log(p_{l,i}/\Delta x^2) ,
\end{equation}
where $N$ is the total number of cells and the index $i$ runs over all the cells. $\Delta x^2$ is the area of the cells, it is added to make the entropy stable against changes of spatial scale as proposed in Ref. \cite{Batty1974}. We take as unit the area our $500 \times 500$ $m^2$ cells and, thus, a change in cell size as those shown in the Appendix for  $1\times 1$ and  $2\times 2$ square kilometers requires a correction factor $4$ and $16$, respectively, as expressed in Equation \eqref{entropy}. The distribution of the population is generally heterogeneous, so $s_{l,c}$ by itself is not telling us anything about characteristic features of the community $l$. To overcome this and also to take into account the finite sampling size, we introduce next a random null model. The $n_{l,c}$ users associated to language $l$ in city $c$ are drawn at random over the city cells according to the total distribution of users to obtain new fractions $p^r_{l,i}$ for language $l$ in each cell $i$, and then we evaluate the following entropy:
\begin{equation}
s_{l,c}^{rand} = - \sum_{i=1}^{N} p^r_{l,i} \, \log(p^r_{l,i}/\Delta x^2) .
\end{equation}
This process is repeated $R$ times to smooth out fluctuations and in this way we obtain an average $\langle s_{l,c}^{rand} \rangle$.  Here, we are interested in the limit of large number of realizations, $R$, in which the users speaking language $l$ would be distributed at random within the local population (fully integrated). The reason to repeat the procedure instead of using in a single run the distribution of the full population is to maintain the effect of the finite number of users speaking $l$. The speakers of this community $l$ can be more or less concentrated in certain areas than the general population. To assess this effect, we define for each city $c$ and detected language $l$ the ratio:
\begin{equation}
\hat{h}_{l,c} = \frac{s_{l,c}}{\langle s_{l,c}^{rand} \rangle} .
\end{equation}
To make the metric further comparable across cities, we further normalized $\hat{h}_{l,c}$ by the value obtained for the local language(s) spoken in city $c$, $\hat{h}_{loc,c}$ (Table S5 in Appendix). If more than one local language is present in the city, the data for all these languages is aggregated to obtain a joint value of $\hat{h}_{loc,c}$. The final definition of the ratio of entropies is thus:
\begin{equation}
h_{l,c} = \hat{h}_{l,c}/\hat{h}_{loc,c} .
\end{equation}
In this way, the information provided takes as baseline the local population and will inform us whether a specific group is spatially segregated or not. According to this definition, low values of $h_{l,c}$ are symptoms of segregation, whereas local languages and those distributed spatially in a similar manner are characterized by $h_{l,c}$ values close to unit. The values of this normalized ratio $h_{l,c}$ constitute the weights of the links in the bipartite network displayed in Figure \ref{fig:fig2}. 

The stability of the spatial entropy in function of different cells sizes (different scales $\Delta x^2$) is studied in the Appendix. We evaluate the relative error among the links of the bipartite network in function of $\Delta x$ taking as reference the unit-like cell with $500$ $m$ side frame. Results are quite stable taking into account the spatial component of entropy related to the side size of the cells of $1000$ and $2000$ meters, respectively, as shown in Figure S6 in Appendix.

\subsection*{Evaluation of the migrant communities spatial distribution's accuracy} 

Twitter has the advantage of being a global source of data, but also the disadvantage of having several uncontrollable biases. Young people are usually over-represented \cite{Mislove2011,Sloan2017}, and most likely the people belonging to the diverse communities are adopting the technology in different ways. If the use of geolocated Twitter is widespread in the host country, this will depend on the maturity of the migrant community: second and third generations are more likely to behave as locals and to adopt generalized technologies in the host population than first generations. On the other hand, things may vary if the technology is already commonly accepted in the country of origin of the community. Certainly, there are communities that are not detected. According to the National Institute for Statistics of Spain (INE, http://www.ine.es), a total of $45,728$ and $54,599$ Chinese citizens are residing in Barcelona and Madrid provinces, respectively, in $2016$. Provinces are territorial divisions that enclose the urban areas and that loosely correspond to the area of analysis taken for our Twitter data. However, the number of users detected tweeting in Chinese is below the threshold of $30$ with a valid residence cell and, therefore, this community does not appear in either of these cities. In the case of this particular group, there may be various reasons for this situation including the relative novelty of Chinese migration to Spain with most of this people belonging to the first generation, as well as the existence of alternatives to Twitter in China such as Sina Weibo. The important question here is thus not whether we find all the communities, but whether we are able to say something meaningful about those detected.

\begin{figure*}
	\centering 
	\includegraphics[width=\linewidth]{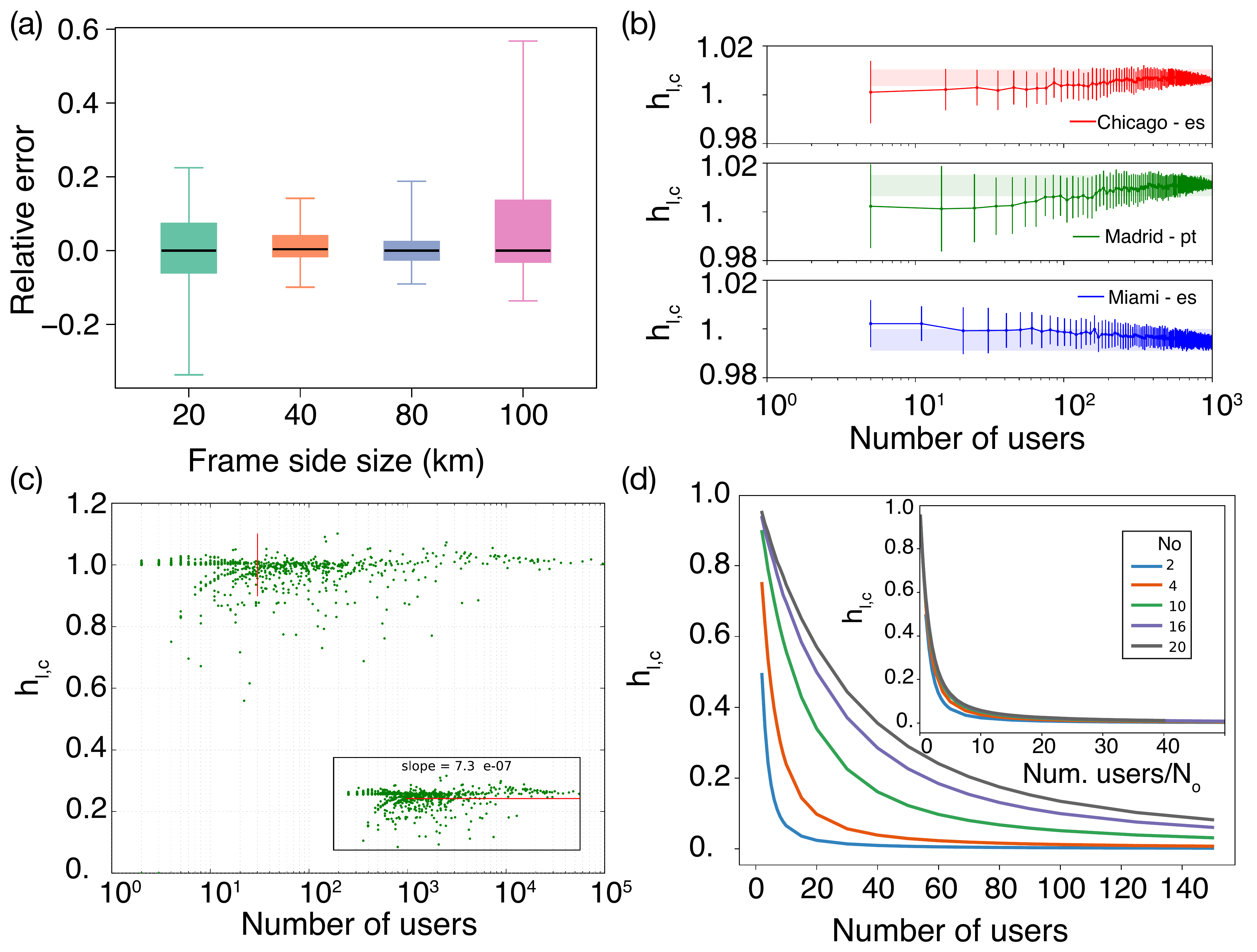}
	\caption{\textbf{Evaluation of the migrant communities spatial distribution.} (a) Box plots of the relative change $\epsilon_{l,c}$ of the link weights in the bipartite spatial integration network taking as reference the $60$ km side frame. (b) The entropy ratio $h_{l,c}$ for three examples of communities with more than $1000$ detected users (Spanish in Chicago and Miami and Portuguese in Madrid). A random sub-sampling is extracted and the calculated ratio of entropies is displayed as a function of the sample size.  (c) The ratio of entropies $h_{l,c}$ as a function of the community size in number of users with a valid residence for all the communities. Every points represent a linguistic community in a city. The red vertical line marks the level of $30$ users taken as a threshold. In the inset, it is shown a zoom-in with the details of the main plot. (d) We present the results concerning the ratio of entropies of a null model in which users belonging to a immigrant community is allowed to reside only in a subset $N_0$ of cells. These users are distributed randomly in the $N_0$ cells, while the local population is randomly distributed across all the gird cells. In the numerical examples, the system contains $100 \times 100 = 10000$ cells. The figure shows how the ratio of entropies changes with the number of users in the immigrant community and how the curves depend in first order on the ratio between the number of users and $N_0$.}
	\label{fig:fig3}
\end{figure*}

Going step by step, let us consider first the influence of the geographical area chosen on the structure of the bipartite network between language communities and cities paying special attention to the weights of its links. For this, recall that we have selected areas of $60 \times 60$ km$^2$ around the barycenter of the $53$ cities considered. These areas have been further divided in cells of $500 \times 500$ m$^2$, which are the basic units of the analysis. The $53$ cities are large megalopolis, still one can wonder if a square frame of $60$ km side is enough to cover all of them, or whether we are including rural areas that could pollute the results. To check the stability of the network  in function of the size of the city boundaries, we evaluate the relative error among the edge weights for different side sizes ($20$, $40$, $80$ and $100$ km) using as reference the original $60 \times 60$ km$^2$ frame. In particular, the relative change $\epsilon_{l,c}$ of the link weights in the bipartite spatial integration network taking as reference the $60$ km side frame is computed as follows,
\begin{equation}
\epsilon_{l,c} = \frac{h_{l,c} - h_{l,c}^{ref}}{h_{l,c}^{ref}} 
\end{equation} 
where $h_{l,c}^{ref}$ represents the edge weight for $60 \times 60$ $km^2$ frame. Box plots displaying the distribution $\epsilon_{l,c}$ values for different frame side sizes can be found in Figure \ref{fig:fig3}a. The network weights are stable for frame side sizes ranging from $40$ to $80$ km. Beyond these values, the differences are increasing, the influence zone is too limited or extended far away from the center into rural areas or other neighboring cities. The value
of $60$ km for the side size is thus a safe choice. It is also worth nothing than the number of detected languages increases with the size of the frame. This number is however quite stable for box sizes ranging from $40$ to $80$ km ($\pm$ 6\% of the reference value). We perform the same analysis over the cell side size, taking as reference the $500$ m side frame. Results are still quite stable increasing the size to $1000$ and $2000$ meters, respectively, as shown in Appendix Figure S7.

A next question to consider concerns the minimum number of users needed to obtain a stable measure of $h_{l,c}$. The number of users for whom we can detect a residence area per community are not very high (Table S6 in Appendix), and in addition we have set a threshold of at least $30$ users to accept the data of a community. Where this value is coming from? To get a first impression of the effect that the user number has on $h_{l,c}$, we select some of the most populous migrant communities, delete a fraction of their users at random and plot in Figure \ref{fig:fig3}b the value of $h_{l,c}$ as a function of the remaining users. Every random extraction produces a different value of $h_{l,c}$, so in the plot we depict the average and the error bars obtained from the standard deviation. Besides, we mark with a shadowed areas the values between which $h_{l,c}$ lies for the extractions with the largest number of users. The results depend on the particular community, but in general the values of  $h_{l,c}$ enter in the shadowed areas between $10$ and $100$ users, $30$ corresponds to the middle ground in logarithmic scale. A more systematic check can be seen in Figure \ref{fig:fig3}c. There, a scatter plot with every value of $h_{l,c}$ for couples language-city is depicted as a function of the number of users associated to the particular community. After $30$ users, there is no more clear dependency between $h_{l,c}$ and the number of users so it must reflect the spatial distribution of the communities. It is also possible to perform a more detailed check in a controlled environment by introducing a null model in which the local population is randomly but uniformly distributed across the grid forming the city, while the immigrant population can only appear in a subset $N_0$ of cells. In those cells the immigrants are also distributed uniformly and randomly. By tuning the number of immigrant users and $N_0$, one can explore how the metric $h_{l,c}$ reacts to finite numbers (see Figure \ref{fig:fig3}d). When the number of immigrants detected is smaller than $N_0$, they are indistinguishable from the local population and thus the ratio $h_{l,c}$ starts in one. As the number of immigrant users gets over $N_0$, the fact that their residence is restricted to a certain area of the city becomes evident and $h_{l,c}$ decays towards a fixed value. As can be seen in the inset of Figure \ref{fig:fig3}d, the main control parameter of the null model is the ratio between the number of immigrant users and $N_0$. The curves showing $h_{l,c}$ as a function of the number of immigrants  collapse by considering them as a function of such ratio. In general terms, the metric $h_{l,c}$ reaches a stable value once the number of immigrants is between $10$ and $20$ times larger than the cells where the community concentrates $N_0$. This model is a worst-case scenario for testing $h_{l,c}$, since the immigrants distribute uniformly while in more realistic applications if a ghetto exists the concentration density will not be uniform. In this latter case, lower number of users are required to measure the stable value of $h_{l,c}$.

\begin{table}
	\centering
	\begin{tabular}{llccc}
		City &Language & $I$ & Z-value&  Autocorrelation\\ \hline\hline
		\multirow{8}{*}{Barcelona} & Total & $0.63$ & $236.5$ & Positive\\
		& Spanish & $0.62$ & $217.0$ & Positive\\  
		& English & $0.50$ & $230.5$ & Positive\\
		& French & $0.37$ & $151.5$  & Positive\\
		& Italian & $0.28$ & $125.8$ & Positive\\
		& Portuguese & $0.32$ & $151.2$ & Positive\\
		& Arabic & $0.08$ & $89.9$ & Random\\
		& East-Slavic & $0.21$ & $112.8$ & Positive\\
		\hline
		
		\multirow{8}{*}{London} & Total & $0.71$ & $66.5$ & Positive\\
		& English & $0.34$ & $35.9$ & Positive\\
		& Spanish & $0.27$ & $28.1$ & Positive\\  
		& French & $0.25$ & $32.1$  & Positive\\
		& Italian & $0.26$ & $31.9$ & Positive\\
		& Portuguese & $0.15$ & $18.5$ & Positive\\
		& Arabic & $0.34$ & $48.5$ & Positive\\
		
		\hline
		\multirow{8}{*}{Madrid} & Total & $0.62$ & $268.6$ & Positive\\
		& Spanish & $0.62$ & $267.3$ & Positive\\  
		& English & $0.32$ & $159.2$ & Positive\\
		& French & $0.37$ & $151.5$  & Positive\\
		& Italian & $0.26$ & $146.3$ & Positive\\
		& Portuguese & $0.44$ & $204.9$ & Positive\\
		& Arabic & $0.07$ & $41.5$ & Random\\
		& East-Slavic & $0.06$ & $37.7$ & Random\\			   
		
		\hline	\hline		   
	\end{tabular}
	\caption{\textbf{Comparison of linguistic communities detection between census and Twitter}. Global Moran's $I$ for a set of common languages detected in Barcelona, London and Madrid. The z-values are calculated after $99$ permutations. The last column refers to the quality and significance of the spatial autocorrelations detected.}
	\label{tab1}
\end{table}

\begin{figure*}
	\centering 
	\includegraphics[width=\linewidth]{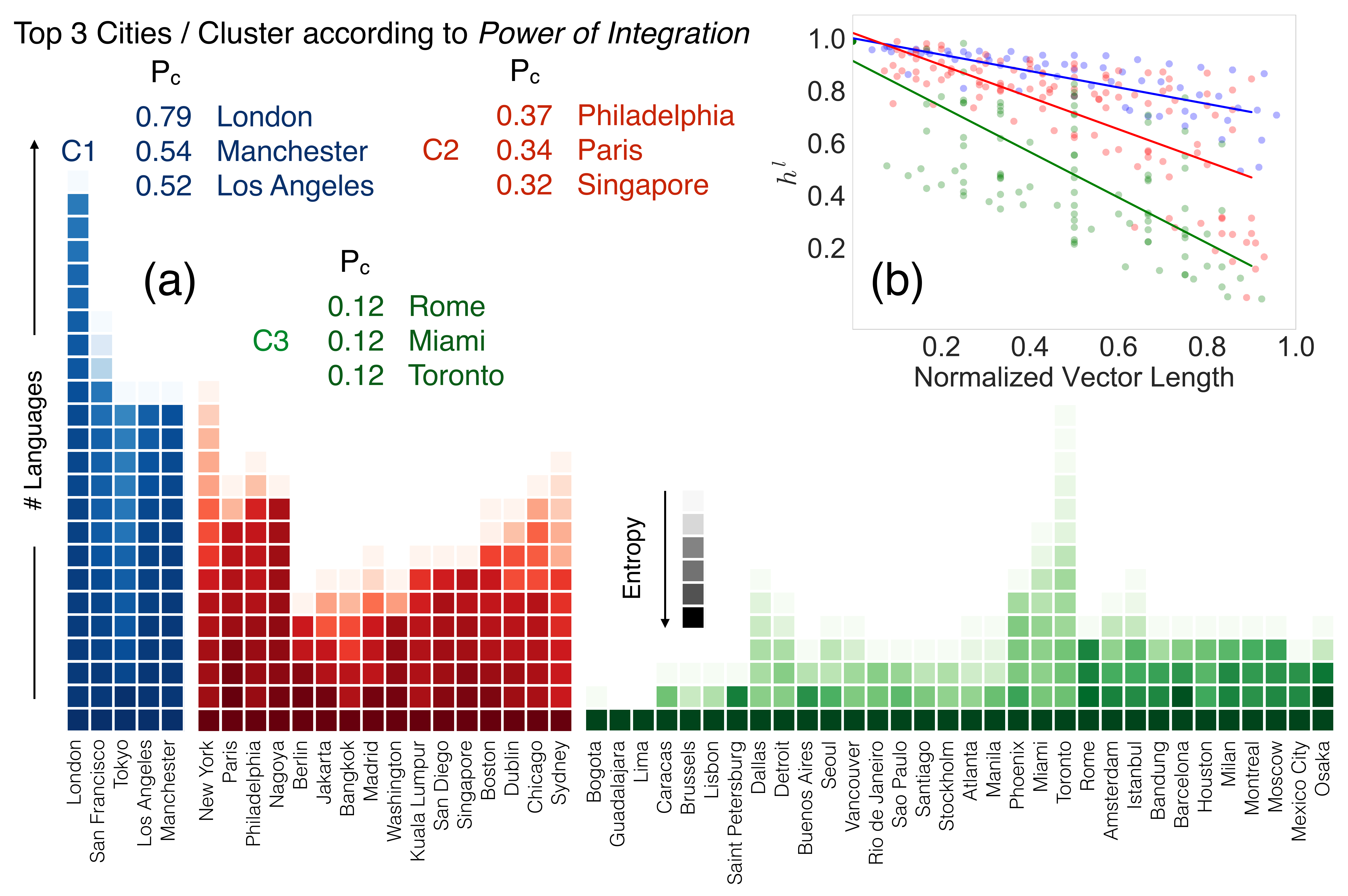}
	\caption{\textbf{Clusters of cities and \textit{Power of Integration}}. In (a), three groups of cities show similar behavior in the number of communities detected and in their levels of integration. The length of the vectors represents the number of languages (communities) detected in each city; the color scale is representative of the decay of the entropy metric; the \textit{Power of Integration} metric lead us to evaluate the potential of each city in uniformly integrating immigrant communities within its own urban area according to entropy values. In (b), decay of $h_{l,c}$ for the cities in each cluster. The points correspond to the values of the elements of $\vec{E_c}$ for each city placed in the x axis according to their index normalized to the total number of languages in $c$. The colors are for cities in the different clusters (C1 blue, C2 red and C3 green), and the lines are minimum square fits to the values of entropy ratios of each cluster.}
	\label{fig:fig4}
\end{figure*}

Finally, we have been also able to run a comparison between the spatial distribution of the communities detected in three cities for which the data from census offices was available. These cities are Barcelona, London and Madrid, and for the comparison we use data from the so-called Continuous Register Statistics in Spain and the Census Office in the UK. In the Spanish case, the information is collected when people residing in a certain area must inform the municipal authorities for tax purposes and to obtain social services such as health care. The smallest spatial units for this dataset are census tracts, so Twitter data must be translated into the same geographical units (see the Appendix for further details).  We employ the Anselin Local Moran's $I$ \cite{Anselin1995} to analyze the level of spatial correspondence of the main migrant communities. This metric provides information on the location, size and spatial coincidence of four types of clusters: a) high-high clusters of significant high values of a variable that are surrounded by high variables of the same variable; b) high-low clusters of significant high values of a variable surrounded by low values of the same variable; c) low-high clusters of significant low values of a variable surrounded by high values of the same variable; and d) low-low clusters of significant low values of a variable surrounded by low values of the same variable. The details are included in Appendix, but a summary with the most important results for a set of linguistic communities common to the three cities are shown in Table \ref{tab1}. The comparison between the location of the residence areas detected with Twitter and those registered in the census is in general good and significant, except for some of the immigrant communities such as Arabic in Barcelona and Madrid or East-Slavic in Madrid where the results lose significance and are compatible with a random distribution.

\subsection*{Power of Integration}

Once the limits of the data and the method to assess the spatial segregation levels of foreign communities have been checked, it is the moment to advance and study what can be said about the way that the cities integrate the foreign groups detected in Twitter. To this end and starting from the bipartite spatial integration network, we perform a clustering analysis based on the distribution of edge weights $h_{l,c}$. For each city $c$, the weights of the edges are sorted in descending order and stored into a vector $\vec{E_c}$. This vector $\vec{E_c}$ contains thus the information on how many foreign linguistic communities have been found in the city $c$ and it quantifies how they are integrated. We can compare next the vectors $\vec{E_c}$ of pairs of cities to assess whether they behave in a similar way respect to the integration of external communities. Similarity metrics usually require the two vectors compared to have the same length. This difficulty can be overcome easily by adding zeros at the end of $\vec{E_c}$ until reaching the maximum length observed in the network  $L_{max}$, namely, for London. We then perform a clustering analysis to find cities exhibiting similar distribution of edge weights by using a k-means algorithm based on Euclidean distances. The results of the analysis are confirmed by repeating the clustering detection with a Hierarchical Clustering Algorithm yielding the same results (see Figure S2 in Appendix). 

Figure \ref{fig:fig4}a shows the three clusters (C1 in blue, C2 in red and C3 in green) obtained after applying the clustering algorithms. These three clusters are characterized by the different rhythm of decay of the entropy values in $\vec{E_c}$ as can be seen in Figure \ref{fig:fig4}b. The first cluster C1 including cities like London, San Francisco, Tokyo or Los Angeles shows the slowest decay. These cities contain in general a number of communities, which are spatially distributed closely mimicking the local population. In the other extreme, the cluster C3 comprises cities with  few or none migrant communities and displaying a high level of spatial segregation for the groups detected. In some cities of this club such as Guadalajara or Lima, we could only detect after applying filters the local languages. However, there are others like Toronto, Miami, Dallas, Rome or Istanbul for which the number of communities is comparable to the cities in the other clusters but the decay of the entropy is way much faster. The communities in their respective $\vec{E_c}$ are highly isolated in comparison with the local population or with similar communities in cities of C1. Finally, there is a middle ground in the cluster C2 containing cities as New York, Paris, Philadelphia, Chicago and Sydney. We introduce a new metric in order to summarize the distribution of entropy and to assess the city's \textit{Power of Integration} (Table S7 in Appendix). This metric is defined, for each city, as: 
\begin{equation}
\displaystyle P_{c} = \frac{L_{c}}{L_{max}} \, Q_{2}\, (1-IQR),
\end{equation}
where $L_{c}$ is the number of languages spoken in city $c$ and $L_{max}$ is the maximum number of languages across the whole set of cities, $Q2$ is the median value of entropy and $IQR$ the interquartile range used as a measure of dispersion.  $P_c$ is maximum when the median of the entropy ratio distribution is one or over,  $IQR = 0$ and the number of languages hosted by city $c$ is the maximum. On the other extreme, it tends to zero when there are no hosted language,  the languages are spatially isolated with $Q_2 = 0$, or when the $IQR = 1$ covering the full range of values.  The top three ranking cities in each cluster according to the Power of Integration are displayed in Figure \ref{fig:fig4}a. According to the full ranking of cities by their \textit{Power of Integration} (Table S7 in Appendix), the metric is able to capture the contribution in the spatial integration process within each urban area: cities belonging to cluster C1 comprises values of $P_{c}$ ranging from Tokyo's $0.41$ to London's $0.79$; the former city shows good integration of massive communities coming from South Korea, Philippines and China. On the other side, the British capital shows almost full spatial mixing of a very large number of foreign communities. Cities belonging to cluster C2 are characterized by values of $P_{c}$ ranging from Jakarta's 0.10 (characterized by mixing segregation behaviors in a scenario of spatial uniformity of most of the communities) to the 0.37 reached on the urban area of Philadelphia; here we found several communities that are uniformly spread within the city, whereas segregation appears focusing on the Arabic speaking community. The cluster as a whole mixes first segregation behaviors in a scenario of several communities involved in the process. Finally, cluster C3 is when both low number of immigrant communities are not well uniformly distributed within the urban areas, proved by the fact that $P_{c}$ are very low. Brussels's 0.01 is due to the low values of entropy of the Turkish community within a scenario of few immigrant communities. Toronto, on the other side, is characterized by a very high number of immigrant communities (comparable to cities found in the cluster C2), not being well spatial integrated within the urban environment. This leads to a $P_{c}$ value of 0.12.  Note that the clusters are obtained directly from the similarity between vectors $\vec{E_c}$ for each city, and later their character is explained by using the decay of the ratios $h_{l,c}$ in the vectors and $P_c$.

\subsection*{Language Integration Network} 

\begin{figure*}
	\includegraphics[width=\linewidth]{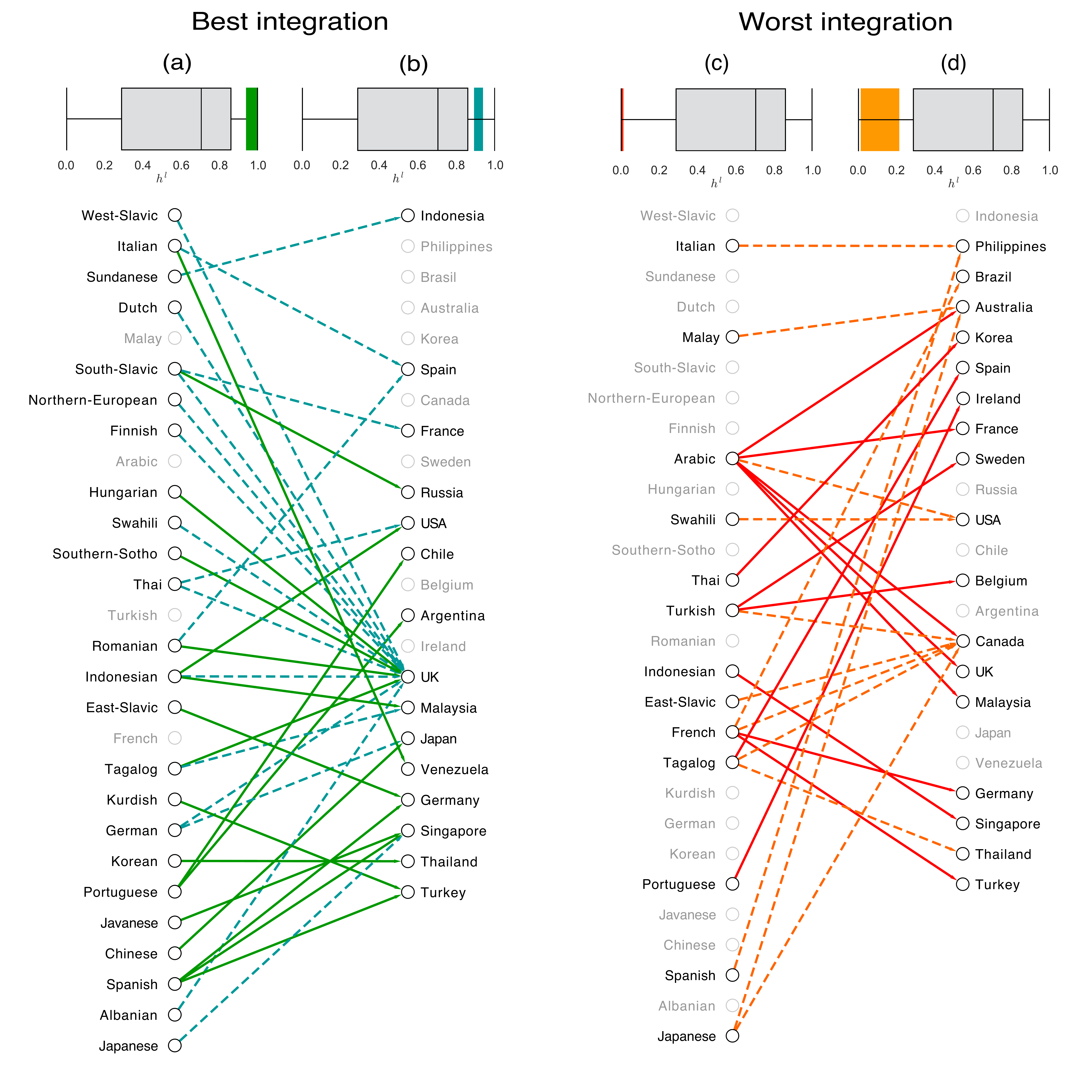}
	\caption{\textbf{Language Integration Network}. We select the sub-network representative of the best levels of spatial integration of languages in countries and display it on the left of the figure. The network is formed by the top $10\%$ links according to the entropy distribution (the spread of the values can be seen in the boxplot (a) in comparison with all the values of $h_{l,c}$). In addition, we include an extra $10\%$ of links  (dash-lines) to the network, those between $10 \%$ and $20 \%$ best links (their spread is in the boxplot (b)). In the network only nodes that belong to the top set are highlighted. Similarly, on the right, the worst levels of spatial integration of languages in countries are shown. We filter out the bottom $10\%$ links according to the entropy distribution (their spread of values is in the boxplot (c)), and add an extra $10\%$ of links to the network (dash-lines), those links between the $10\%$ and $20\%$ worst in the ranking. Their spread is in the boxplot (d). As before, only the nodes that belong to the worst set are highlighted.}
	\label{fig:fig5}
\end{figure*}

The bipartite spatial integration network can be also be projected into the language side to gain insights on the level of integration of languages into the different countries (see Table S8 in Appendix). We do the analysis at the country level because we assume that the integration of the immigrant communities is similar across the cities of the same country. When there are more than one city in the country, we take the average value of the entropy $h_{l,c}$ to build the network. The best and the worst cases of integration are displayed in  Figure \ref{fig:fig5} left and right. Before proceeding to the analysis, it is important to mention that English has been excluded from the network because of its role as \enquote{lingua franca} \cite{Ronen2014}. Moreover, the role of English is dominant mainly in the worst links in terms of integration (see Figure S3 in Appendix for more details). We select two thresholds of levels of integration of language in countries: in the top set (Figure \ref{fig:fig5} left) the strong \textit{Power of Integration} of UK cities (London and Manchester) sets its dominant role in uniformly spatial integrating several communities. Several patterns of uniform spatial integration appear, such as the Italian community in Venezuela, and the Spanish-speaking in Germany, Singapore and Turkey; the latter country shows uniformly distributed communities of Spanish people (due to historical migrations of Spanish Jews dating as far back as the 15th century), and Kurdish (largest ethnic minority in Istanbul). South-Slavic and East-Slavic communities keep their traditional presence in Russia and Germany. Increasing the threshold of the link weights, UK leads in the role of hosting diverse communities and some other patterns emerge, such as the German presence in Japan and UK. By contrast (Figure \ref{fig:fig5} right), Arabic rises as the most common spatially segregated community followed by French-speaking communities that appear to be spatially concentrated in other European countries such as Germany and Turkey. Increasing the threshold further, results in more forms of segregation appearing in Canada (East-Slavic, French and Tagalog), Australia (Malay and Japanese), Brazil (French) and Philippines (Italian and Spanish). Note that the segregation can occur on the two extremes of the economic spectrum: poor people may need to live in ghetto-like areas but also wealthier communities  may  concentrate with respect to the general local population as it seems to be the case for Italian and Spanish speaking minorities in the Philippines or the English speaking community in Rome. 

\section*{Discussion}

People are constantly moving within cities and countries, looking for jobs, experiences or just for better life conditions, facing the fact of the integration in habits and laws of new local cultures. Migration flows have been studied so far by means of surveys and census data that cover from the number of people living outside their country of birth to place of residency to features of the labor market. However, census and surveys have the disadvantage of a very high cost, geographical limitations and, typically, they have slow update frequencies. Recent works by experts in the area highlight the dare need of more agile data sources about mobility and settlement patterns of immigrant and refugee communities.  

Rather than using these classical sources, in this work we explore the capability of the online social networks to provide information about the integration of immigrant communities. In particular, we use Twitter to connect users to their residence place and via a language \enquote{algebra} to determine their cultural background. This allows us to study how spatial and linguistic characteristics of people vary within the cities they are living in, and how the cities spatially integrate the diversity of languages and cultures characteristic of the global metropolises today. It is necessary to admit the potential biases of the data: the social network penetration through socio-economic hierarchies, age, generations and countries is different. This is precisely the reason why we do not detect all the possible communities in the cites under consideration. Still we have introduced a method compressing a metric that is not so sensitive to the small numbers in the users detected. As can be seen, in the validation exercise the results in the cities where we can compare with the census are significant for communities with more than $30$ users. This method in general measures how well different communities are spatially integrated/segregated within urban areas. Our findings provide a new way to observe the patterns of historically  immigration of people to urban areas, and any potential changes that might arise in the areas of residence. We are able to move beyond the estimation of past, current and foreshadowed global flows toward a better comprehension of the integration phenomena on a city scale. Residents' online communications can thus let us assess in an indirect way if the cultural background has been kept inside communities, although impacted on different levels by local welcoming and hosting policies. This method provides an extra alternative to the toolkit of researchers in sociology and urbanism as well as direct view in close to real time on the potential problems of integration that may appear in different areas of the cities, a knowledge that can be of great value to public managers.

\section*{Acknowledgments}

Partial financial support has been received from the Spanish Ministry of Economy (MINECO) and FEDER (EU) under the project ESOTECOS (FIS2015-63628-C2-2-R), and from the EU Commission through project INSIGHT. The work of M-HS-O was supported in part by a post-doctoral fellowship of MINECO at Universidad Complutense de Madrid (FPDI 2013/17001). BG thanks the Moore and Sloan Foundations for support as part of the Moore-Sloan Data Science Environment at New York University.

\section*{Author contributions}

F.L., M.L., G.R. and J.J.R. designed and performed research; F.L., M.L., M.H.S.O. and B.G. retrieved data; F.L., M.L., M.H.S.O. and B.G. analyzed data; and F.L., M.L., M.H.S.O., G.R., B.G. and J.J.R. wrote the paper.

\bibliographystyle{unsrt}
\bibliography{Immigrants.bib}

\newpage
\vspace*{2cm}
\newpage
\onecolumngrid

\makeatletter
\renewcommand{\fnum@figure}{\sf\textbf{\figurename~\textbf{S}\textbf{\thefigure}}}
\renewcommand{\fnum@table}{\sf\textbf{\tablename~\textbf{S}\textbf{\thetable}}}
\makeatother

\setcounter{figure}{0}
\setcounter{table}{0}
\setcounter{equation}{0}

\section*{Appendix}

\begin{figure}[!h]
\centering
\includegraphics[width=12cm]{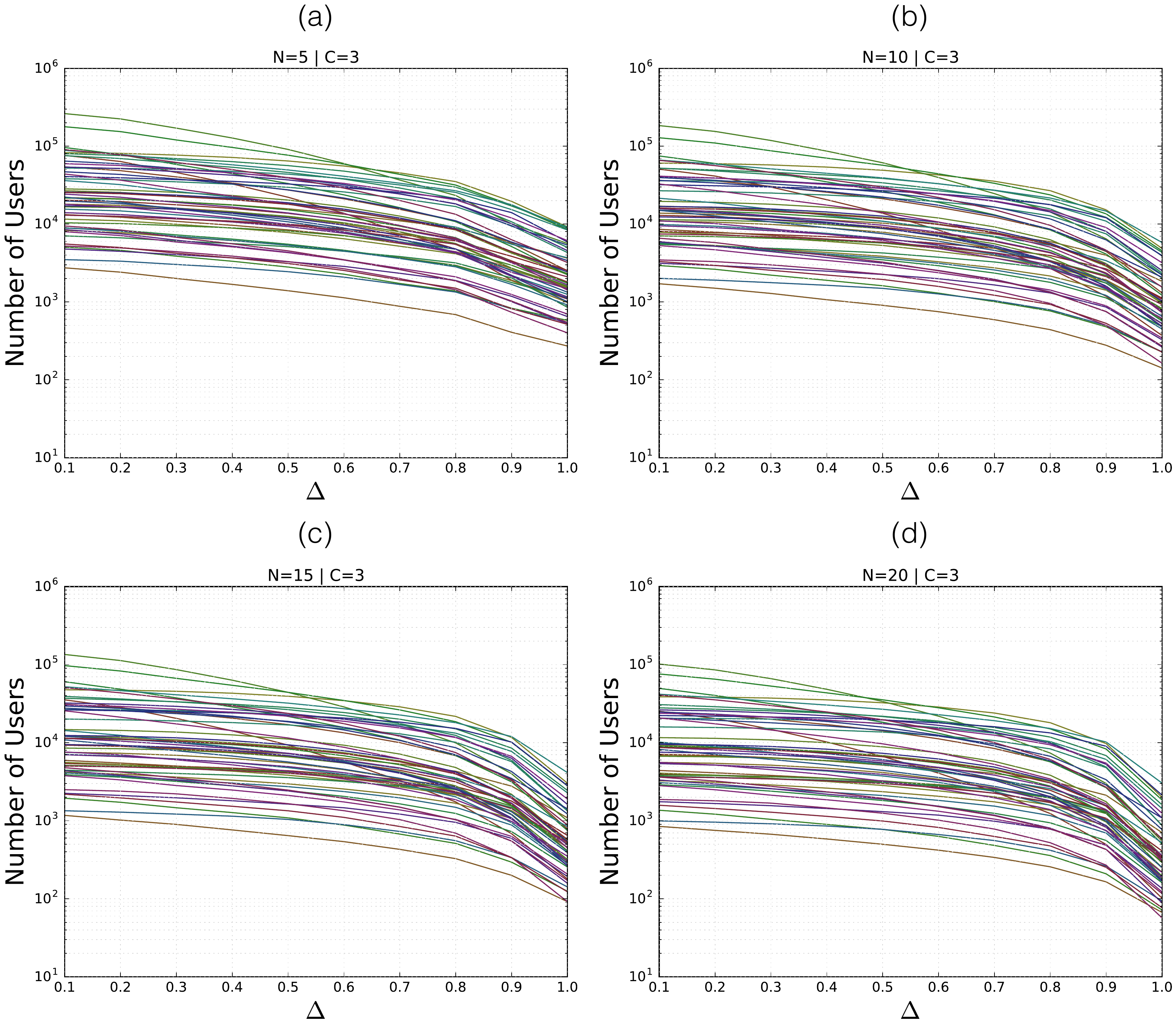}
\caption{\textbf{Number of reliable users as a function of \textit{N} and $\Delta$}. Each line represents the trend of each city in the number of users according to the ratio between \textit{N} and the total number of hours of activity for each user ($\Delta$). Set as $C$=3 the number of months for consecutive activities, (a) refers to $N$=5, (b) to $N$=10, (c) to $N$=15 and (d) to $N$=20.}
\label{FigS1}
\end{figure}

\begin{figure}[ht!]
\centering
\includegraphics[width=14cm]{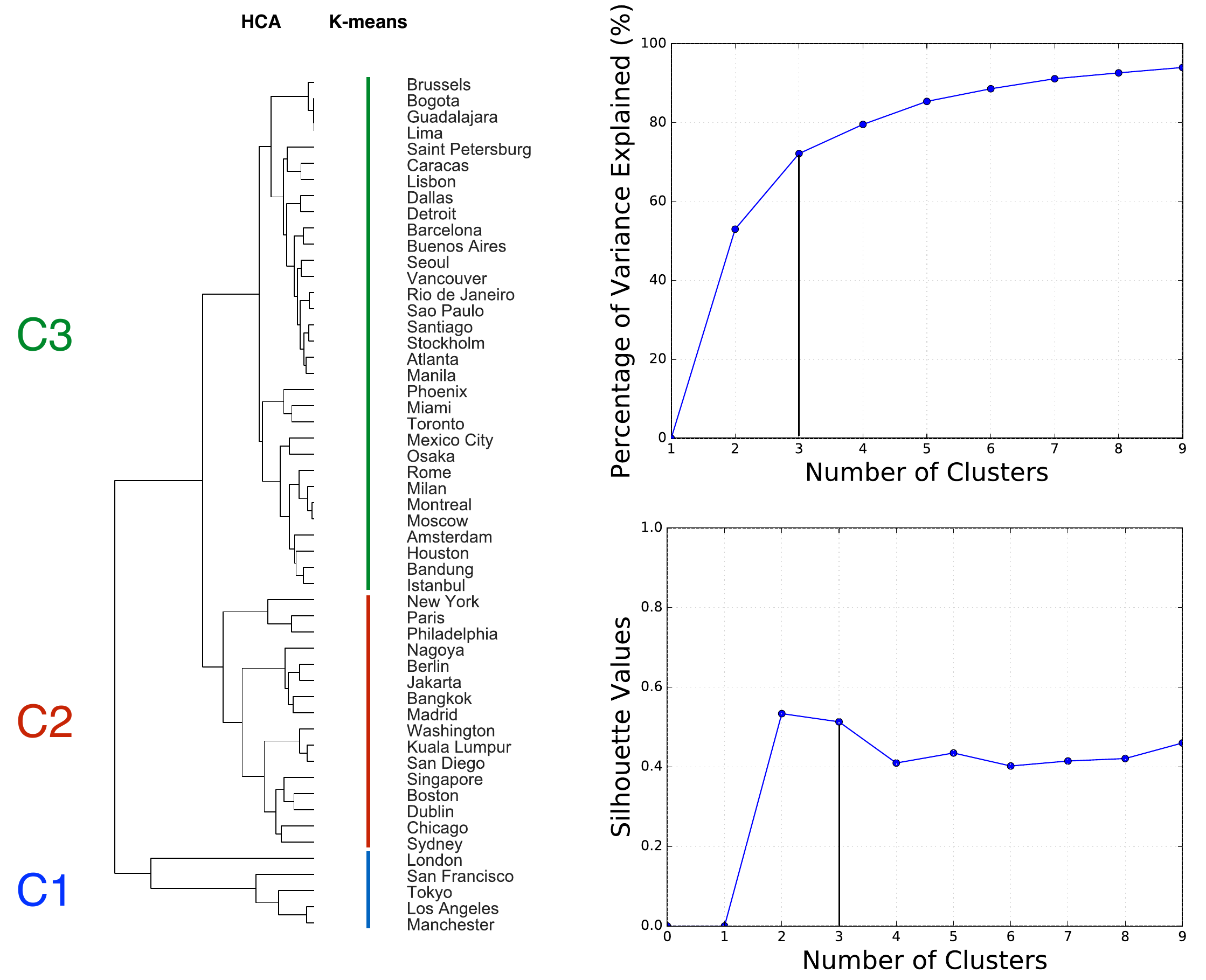}
\caption{\textbf{Comparison of a k-means and hierarchical clustering algorithms over the vectors of the Bipartite Spatial Integration Network}. C1, C2 and C3 are the clusters obtained through both algorithms, choosing as 3 the initial number of clusters to assign to the k-means analysis.} 
\label{FigS2}
\end{figure}

\begin{figure}
  \centering 
  \includegraphics[width=14cm]{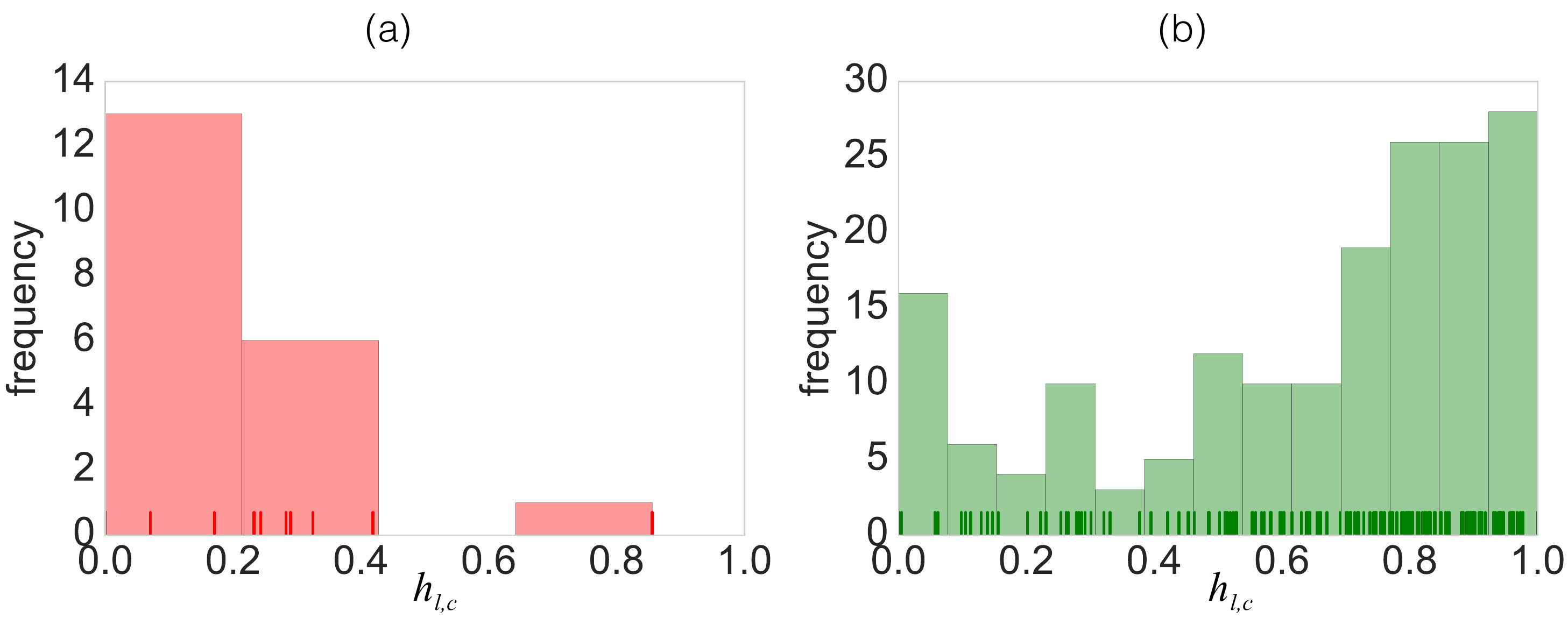}
  \caption{\textbf{Distribution of degree and weights in \textit{H} with and without English}. Distribution of weights for the full network (a) and in the network obtained removing English from the nodes of the Bipartite Spatial Integration Network (b). As shown, English is dominant in the worst links in terms of spatial integration.}
\label{FigS3}
\end{figure}

\clearpage
\section*{Supporting Tables}

\begin{table}[h]
\centering
\scalebox{1}{
\begin{tabular}{l l l}

City & tweets & users\\ \hline

Amsterdam & 2082040 & 175107 \\
Atlanta & 4896616 & 250803 \\
Bandung & 7980207 & 487209 \\
Bangkok & 9389888 & 265672 \\
Barcelona & 2249807 & 168189 \\
Berlin & 703551 & 64360 \\
Bogota & 3370596 & 195830 \\
Boston & 3500542 & 206385 \\
Brussels & 1434065 & 84140 \\
Buenos Aires & 18421250 & 459534 \\
Caracas & 1376503 & 108769 \\
Chicago & 5008746 & 304844 \\
Dallas & 6650105 & 253760 \\
Detroit & 3885662 & 144775 \\
Jakarta & 28304891 & 1260903 \\
Dublin & 1516170 & 89259 \\
Guadalajara & 704530 & 62403 \\
Houston & 5337061 & 193876 \\
Istanbul & 19438021 & 838347 \\
Kuala Lumpur & 16730001 & 412560 \\
Lima & 1356562 & 94234 \\
Lisbon & 3341088 & 57201 \\
London & 11167058 & 698424 \\
Los Angeles & 12458292 & 600476 \\
Madrid & 5605452 & 275857 \\
Manchester & 6940211 & 337083 \\
Manila & 19453573 & 449308 \\

\end{tabular}
\hspace*{2 cm}
\begin{tabular}{l l l}

City & tweets & users\\ \hline

Mexico City & 4458228 & 322462 \\
Miami & 3716735 & 241676 \\
Milan & 1493614 & 103383 \\
Montreal & 631844 & 52851 \\
Moscow & 2805144 & 141589 \\
Nagoya & 3941030 & 162606 \\
New York & 12960258 & 734100 \\
Osaka & 10607046 & 351628 \\
Paris & 10929091 & 335553 \\
Philadelphia & 5841765 & 247875 \\
Phoenix & 2970897 & 136501 \\
Rio de Janeiro & 20907590 & 295565 \\
Rome & 1240080 & 96213 \\
Saint Petersburg & 1124887 & 63828 \\
San Diego & 2267882 & 164061 \\
San Francisco & 4615005 & 284215 \\
Santiago & 3022229 & 148786 \\
Sao Paulo & 18744302 & 371032 \\
Seoul & 1338750 & 118509 \\
Singapore & 7530611 & 292571 \\
Stockholm & 784807 & 51729 \\
Sydney & 1226072 & 80349 \\
Tokyo & 15029229 & 644683 \\
Toronto & 2281968 & 150786 \\
Vancouver & 660837 & 58984 \\
Washington & 6515118 & 330808 \\

\\
\end{tabular}
}
\caption{\textbf{Number of tweets and users detected in each city after filtering out bots and multi-user accounts.}}
\label{TabS1}
\end{table}

\begin{table}[h]
\centering
\scalebox{1}{
\begin{tabular}{l l l}

City & Latitude & Longitude\\ \hline

Amsterdam&52.370216&4.895168 \\
Atlanta&33.748995&-84.387982 \\
Bandung&-6.914744&107.609811 \\
Bangkok&13.727896&100.524123 \\
Barcelona&41.385064&2.173403 \\
Berlin&52.519171&13.406091 \\
Bogota&4.598056&-74.075833 \\
Boston&42.358431&-71.059773 \\
Brussels&50.85034&4.35171 \\
Buenos Aires&-34.603723&-58.381593 \\
Caracas&10.491016&-66.902061 \\
Chicago&41.878114&-87.629798 \\
Dallas&32.78014&-96.800451 \\
Detroit&42.331427&-83.045754 \\
Jakarta&-6.211544&106.845172 \\
Dublin&53.349805&-6.26031 \\
Guadalajara&20.67359&-103.343803 \\
Houston&29.760193&-95.36939 \\
Istanbul&41.00527&28.97696 \\
Kuala Lumpur&3.139003&101.686855 \\
Lima&-12.047816&-77.062203 \\
Lisbon&38.725299&-9.150036 \\
London&51.511214&-0.119824 \\
Los Angeles&33.95&-118.14 \\
Madrid&40.416775&-3.70379 \\
Manchester&53.479324&-2.248485 \\
Manila&14.599512&120.984219 \\

\end{tabular}
\hspace*{2 cm}
\begin{tabular}{l l l}

City & Latitude & Longitude\\ \hline

Mexico City&19.42705&-99.127571 \\
Miami&25.788969&-80.226439 \\
Milan&45.465454&9.186516 \\
Montreal&45.50867&-73.553992 \\
Moscow&55.755826&37.6173 \\
Nagoya&35.181446&136.906398 \\
New York&40.714353&-74.005973 \\
Osaka&34.693738&135.502165 \\
Paris&48.856614&2.352222 \\
Philadelphia&39.952335&-75.163789 \\
Phoenix&33.448377&-112.074037 \\
Rio de Janeiro&-22.903539&-43.209587 \\
Rome&41.892916&12.48252 \\
Saint Petersburg&59.93428&30.335099 \\
San Diego&32.715329&-117.157255 \\
San Francisco&37.774929&-122.419416 \\
Santiago&-33.46912&-70.641997 \\
Sao Paulo&-23.548943&-46.638818 \\
Seoul&37.566535&126.977969 \\
Singapore&1.352083&103.819836 \\
Stockholm&59.32893&18.06491 \\
Sydney&-33.867487&151.20699 \\
Tokyo&35.689487&139.691706 \\
Toronto&43.653226&-79.383184 \\
Vancouver&49.261226&-123.113927 \\
Washington&38.907231&-77.036464 \\
\\
\end{tabular}
}
\caption{\textbf{Coordinates of the centers of the frame for each city.}}
\label{TabS2}
\end{table}

\begin{table}[h]
\centering
\scalebox{1}{
\begin{tabular}{l r}

City & Resident Users\\ \hline
Amsterdam        &    4986 \\
Atlanta          &    8474 \\
Bandung          &   27818 \\
Bangkok          &   25659 \\
Barcelona        &    9957 \\
Berlin           &    1301 \\
Bogota           &   19353 \\
Boston           &    8989 \\
Brussels        &    2325 \\
Buenos Aires      &   48934 \\
Caracas          &    4613 \\
Chicago          &   15397 \\
Dallas           &   15549 \\
Detroit          &   10652 \\
Jakarta         &   98997 \\
Dublin           &    5480 \\
Guadalajara      &    3459 \\
Houston          &   11413 \\
Istanbul         &  101556 \\
Kuala Lumpur      &   41084 \\
Lima             &    2003 \\
Lisbon           &    4321 \\
London           &   37402 \\
Los Angeles       &   70592 \\
Madrid           &   15447 \\
Manchester       &   23836 \\
Manila           &   20093 \\

\end{tabular}
\hspace*{2 cm}
\begin{tabular}{l r}

City & Resident Users\\ \hline
Mexico City       &   18079 \\
Miami            &    7754 \\
Milan           &    4243 \\
Montreal         &    2613 \\
Moscow           &    9673 \\
Nagoya           &    7589 \\
NewYork          &   34325 \\
Osaka            &   16348 \\
Paris            &   19757 \\
Philadelphia     &   13679 \\
Phoenix          &    9259 \\
Rio de Janeiro     &   37177 \\
Rome             &    1994 \\
Saint Petersburg &    3819 \\
San Diego         &    5014 \\
San Francisco     &   25504 \\
Santiago         &   10066 \\
Sao Paulo         &   21862 \\
Seoul            &    3099 \\
Singapore        &   20997 \\
Stockholm        &    2668 \\
Sydney           &    4751 \\
Tokyo            &   75929 \\
Toronto          &    8737 \\
Vancouver        &    2298 \\
Washington       &   10147 \\
\\
\end{tabular}
}
\caption{\textbf{Total number of residents users detected in the cities.}}
\label{TabS3}
\end{table}

\begin{table}
\centering
\scalebox{1}{
\begin{tabular}{l r}

Detected language & Aggregated group\\ \hline
Albanian&Albanian \\
Arabic&Arabic \\
Belarusian&\textbf{East Slavic} \\
Bosnian&\textbf{South Slavic} \\
Bulgarian&\textbf{South Slavic} \\
Catalan&Catalan\\
Chinese&Chinese\\
Croatian&\textbf{South Slavic}\\
Czech&\textbf{West Slavic}\\
Danish&\textbf{Northern European}\\
Dutch&Dutch (including Flemish)\\
English&English\\
Faroese&\textbf{Northern European}\\
Finnish&Finnish\\
French&French\\
German&German\\
Greek&Greek\\
Haitian&Haitian\\
Hungarian&Hungarian\\
Icelandic&\textbf{Northern European}\\
Indonesian&Indonesian\\
Irish&Irish\\
Italian&Italian\\
Japanese&Japanese\\
Javanese&Javanese\\
Korean&Korean\\

\end{tabular}
\hspace*{2 cm}
\begin{tabular}{l r}

Detected Language & Aggregated Group\\ \hline
Kurdish&Kurdish\\
Lettonian&\textbf{Baltic}\\
Lituanian&\textbf{Baltic}\\
Macedonian&\textbf{South Slavic}\\
Malay&Malay\\
Norwegian&\textbf{Northern European}\\
Polish&\textbf{West Slavic}\\
Portuguese&Portuguese\\
Romanian&Romanian\\
Russian&\textbf{East Slavic}\\
Serbian&\textbf{South Slavic}\\
Serbo-Croatian&\textbf{South Slavic}\\
Slovak&\textbf{West Slavic}\\
Slovenian&\textbf{South Slavic}\\
Southern Sotho&Southern Sotho\\
Spanish&Spanish\\
Swahili&Swahili\\
Swedish&\textbf{Northern European}\\
Sundanese&Sundanese\\
Tagalog&Tagalog\\
Thai&Thai\\
Turkish&Turkish\\
Ukrainian&\textbf{East Slavic}\\
Vietnamese&Vietnamese\\
\\
\\
\end{tabular}
}
\caption{\textbf{Language aggregation process}. A main Aggregated group has been associated to each language detected in the framework, to overlap "mutually intelligible" issues in the detection.}
\label{TabS4}
\end{table}

\begin{table}[h]
\centering
\scalebox{1}{
\begin{tabular}{lr}

City & Local Culture\\ \hline
Amsterdam & Dutch \\
Atlanta & English \\
Bandung & Indonesian \\
Bangkok & Thai \\
Barcelona & Spanish/Catalan \\
Berlin & German \\
Bogota & Spanish \\
Boston & English \\
Brussels & French/Flemish \\
Buenos Aires & Spanish \\
Caracas & Spanish \\
Chicago & English \\
Dallas & English \\
Detroit & English \\
Jakarta & Indonesian \\
Dublin & English/Irish \\
Guadalajara & Spanish \\
Houston & English \\
Istanbul & Turkish \\
Kuala Lumpur & Malay \\
Lima & Spanish \\
Lisbon & Portuguese \\
London & English \\
Los Angeles & English \\
Madrid & Spanish \\
Manchester & English \\
Manila & Tagalog \\

\end{tabular}
\hspace*{2 cm}
\begin{tabular}{l r}

City & Local Culture\\ \hline
Mexico City & Spanish \\
Miami & English \\
Milan & Italian \\
Montreal & French/English \\
Moscow & East-Slavic \\
Nagoya & Japanese \\
New York & English \\
Osaka & Japanese \\
Paris & French \\
Philadelphia & English \\
Phoenix & English \\
Rio de Janeiro & Portuguese \\
Rome & Italian \\
Saint Petersburg & East-Slavic \\
San Diego & English \\
San Francisco & English \\
Santiago & Spanish \\
Sao Paulo & Portuguese \\
Seoul & Korean \\
Singapore & Malay/Chinese/English/Tamil \\
Stockholm &  Northern-European \\
Sydney & English \\
Tokyo & Japanese \\
Toronto & English \\
Vancouver & English \\
Washington & English \\
\\
\end{tabular}
}
\caption{\textbf{Cities and local languages}. Each city has been associated to its main local language; Barcelona, Brussels, Dublin, Montreal and Singapore have been related to more than one language due to the coexistence of multiple languages in the same urban area.}
\label{TabS5}
\end{table}

\begin{table}[!p]
\centering
\includegraphics[width=\linewidth]{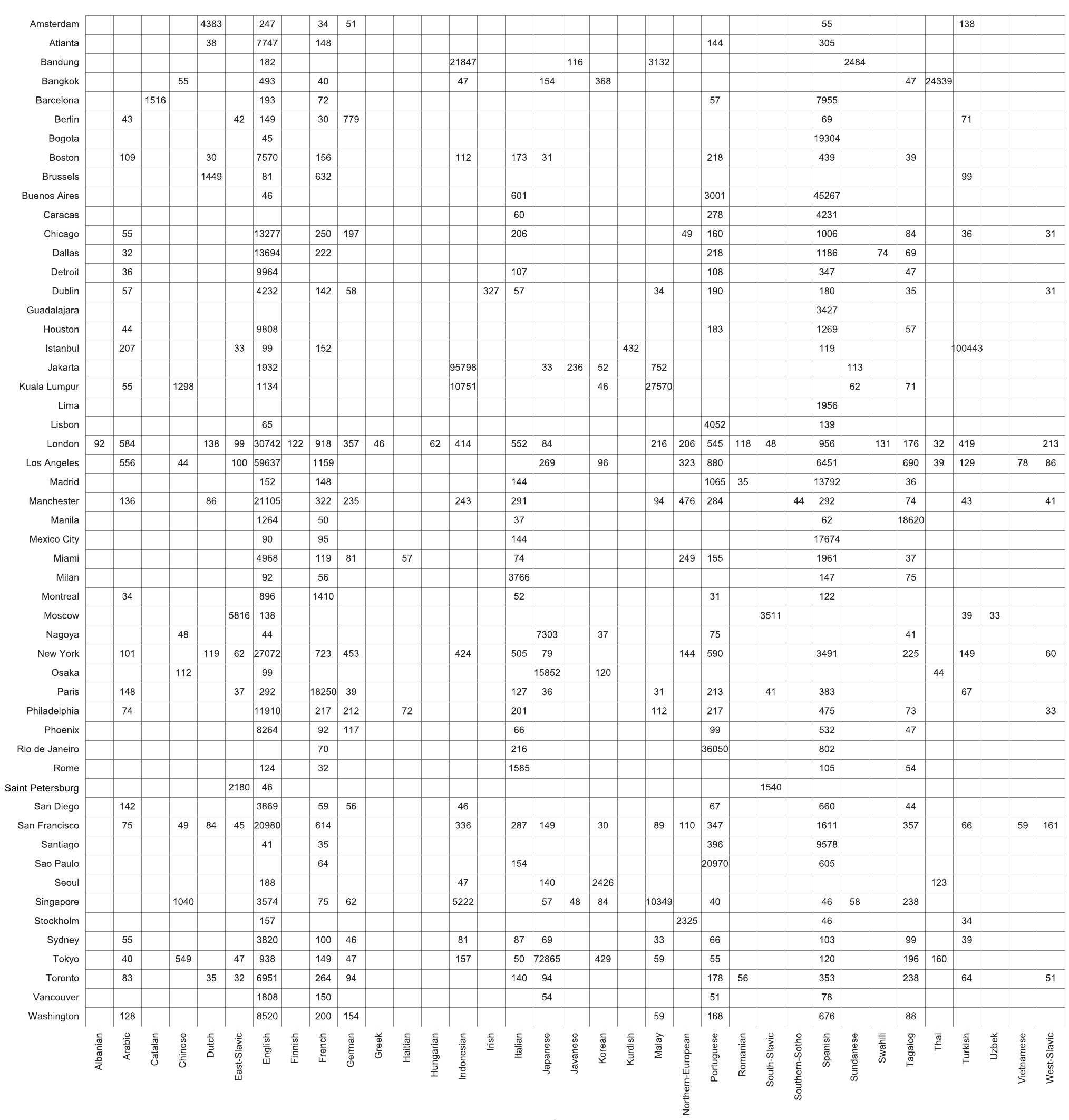}
\caption{\textbf{Number of resident users per language and per city.}}
\label{TabS6}
\end{table}

\begin{table}[!h]
\centering
\scalebox{0.8}{
\begin{tabular}{lllllll}

Cluster & City & Q1 & Q2 & Q3 & IQR & $P_{c}$ \\ \hline
C1 & London & 0.81 & 0.91 & 0.95 & 0.13 & 0.789 \\
C1 & Manchester & 0.91 & 0.95 & 0.96 & 0.06 & 0.543 \\
C1 & Los Angeles & 0.87 & 0.93 & 0.96 & 0.09 & 0.518 \\
C1 & San Francisco & 0.77 & 0.83 & 0.92 & 0.15 & 0.522 \\
C1 & Tokyo & 0.71 & 0.80 & 0.87 & 0.16 & 0.413 \\
C2 & Philadelphia & 0.88 & 0.90 & 0.92 & 0.04 & 0.375 \\
C2 & Paris & 0.76 & 0.81 & 0.90 & 0.14 & 0.336 \\
C2 & Singapore & 0.81 & 0.86 & 0.95 & 0.15 & 0.319 \\
C2 & New York & 0.31 & 0.64 & 0.85 & 0.54 & 0.180 \\
C2 & Kuala Lumpur & 0.83 & 0.87 & 0.90 & 0.07 & 0.246 \\
C2 & San Diego & 0.82 & 0.88 & 0.93 & 0.12 & 0.236 \\
C2 & Boston & 0.65 & 0.80 & 0.88 & 0.23 & 0.241 \\
C2 & Chicago & 0.53 & 0.82 & 0.84 & 0.31 & 0.247 \\
C2 & Dublin & 0.57 & 0.79 & 0.87 & 0.29 & 0.220 \\
C2 & Sydney & 0.27 & 0.65 & 0.75 & 0.48 & 0.161 \\
C2 & Washington & 0.72 & 0.82 & 0.84 & 0.13 & 0.217 \\
C2 & Madrid & 0.61 & 0.91 & 0.94 & 0.33 & 0.159 \\
C2 & Nagoya & 0.76 & 0.86 & 0.97 & 0.22 & 0.146 \\
C2 & Bangkok & 0.40 & 0.77 & 0.84 & 0.43 & 0.133 \\
C2 & Berlin & 0.44 & 0.77 & 0.90 & 0.46 & 0.108 \\
C2 & Jakarta & 0.42 & 0.63 & 0.82 & 0.40 & 0.099 \\
C3 & Amsterdam & 0.30 & 0.52 & 0.74 & 0.44 & 0.063 \\
C3 & Atlanta & 0.10 & 0.21 & 0.41 & 0.31 & 0.026 \\
C3 & Bandung & 0.36 & 0.66 & 0.77 & 0.41 & 0.068 \\
C3 & Barcelona & 0.36 & 0.60 & 0.80 & 0.44 & 0.044 \\
C3 & Bogota & 0.25 & 0.50 & 0.75 & 0.50 & 0.011 \\
C3 & Brussels & 0.12 & 0.24 & 0.62 & 0.50 & 0.011 \\

\end{tabular}
\hspace*{0.5 cm}
\begin{tabular}{lllllll}

Cluster & City & Q1 & Q2 & Q3 & IQR & $P_{c}$ \\ \hline
C3 & Buenos Aires & 0.23 & 0.51 & 0.80 & 0.57 & 0.029 \\
C3 & Caracas & 0.25 & 0.50 & 0.75 & 0.50 & 0.022 \\
C3 & Dallas & 0.19 & 0.34 & 0.40 & 0.20 & 0.071 \\
C3 & Detroit & 0.19 & 0.39 & 0.45 & 0.26 & 0.064 \\
C3 & Guadalajara & 1.00 & 1.00 & 1.00 & 0.00 & 0.000 \\
C3 & Houston & 0.51 & 0.57 & 0.63 & 0.12 & 0.087 \\
C3 & Istanbul & 0.16 & 0.57 & 0.69 & 0.52 & 0.071 \\
C3 & Lima & 1.00 & 1.00 & 1.00 & 0.00 & 0.000 \\
C3 & Lisbon & 0.16 & 0.32 & 0.66 & 0.50 & 0.014 \\
C3 & Manila & 0.11 & 0.22 & 0.53 & 0.41 & 0.023 \\
C3 & Mexico City & 0.54 & 0.74 & 0.82 & 0.28 & 0.070 \\
C3 & Miami & 0.27 & 0.41 & 0.43 & 0.16 & 0.121 \\
C3 & Milan & 0.57 & 0.76 & 0.78 & 0.21 & 0.103 \\
C3 & Montreal & 0.61 & 0.69 & 0.72 & 0.11 & 0.107 \\
C3 & Moscow & 0.66 & 0.72 & 0.76 & 0.10 & 0.113 \\
C3 & Osaka & 0.25 & 0.83 & 0.99 & 0.75 & 0.037 \\
C3 & Phoenix & 0.41 & 0.47 & 0.56 & 0.15 & 0.105 \\
C3 & Rio de Janeiro & 0.33 & 0.47 & 0.63 & 0.29 & 0.044 \\
C3 & Rome & 0.78 & 0.79 & 0.88 & 0.10 & 0.124 \\
C3 & Saint Petersburg & 0.40 & 0.80 & 0.90 & 0.50 & 0.035 \\
C3 & Santiago & 0.21 & 0.38 & 0.61 & 0.40 & 0.030 \\
C3 & Sao Paulo & 0.30 & 0.48 & 0.67 & 0.37 & 0.040 \\
C3 & Seoul & 0.28 & 0.29 & 0.60 & 0.32 & 0.034 \\
C3 & Stockholm & 0.25 & 0.37 & 0.56 & 0.32 & 0.033 \\
C3 & Toronto & 0.10 & 0.33 & 0.47 & 0.37 & 0.117 \\
C3 & Vancouver & 0.19 & 0.38 & 0.46 & 0.28 & 0.047 \\
\\
\end{tabular}
}
\caption{\textbf{Power of Integration of Cities.}}
\label{TabS7}
\end{table}

\begin{table}[!ht]
\centering
\scalebox{1}{
\begin{tabular}{lr}

City & Local Culture\\ \hline
Amsterdam & Netherlands \\
Atlanta & USA \\
Bandung & Indonesia \\
Bangkok & Thailand \\
Barcelona & Spain \\
Berlin & Germany \\
Bogota & Colombia \\
Boston & USA \\
Brussels & Belgium \\
Buenos Aires & Argentina \\
Caracas & Venezuela \\
Chicago & USA \\
Dallas & USA \\
Detroit & USA \\
Jakarta & Indonesia \\
Dublin & Ireland \\
Guadalajara & Mexico \\
Houston & USA \\
Istanbul & Turkey \\
Kuala Lumpur & Malaysia \\
Lima & Perù \\
Lisbon & Portugal \\
London & UK \\
Los Angeles & USA \\
Madrid & Spain \\
Manchester & UK \\
Manila & Philippines \\

\end{tabular}
\hspace*{2 cm}
\begin{tabular}{l r}

City & Local Culture\\ \hline
Mexico City & Mexico \\
Miami & USA \\
Milan & Italy \\
Montreal & Canada \\
Moscow & Russia \\
Nagoya & Japan \\
New York & USA \\
Osaka & Japan \\
Paris & France \\
Philadelphia & USA \\
Phoenix & USA \\
Rio de Janeiro & Brazil \\
Rome & Italy \\
Saint Petersburg & Russia \\
San Diego & USA \\
San Francisco & USA \\
Santiago & Chile \\
Sao Paulo & Brazil \\
Seoul & Korea \\
Singapore & Singapore \\
Stockholm & Sweden \\
Sydney & Australia \\
Tokyo & Japan \\
Toronto & Canada \\
Vancouver & Canada \\
Washington & USA \\
\\
\end{tabular}
}
\caption{\textbf{City/Country Correspondence}.}
\label{TabS8}
\end{table}

\clearpage
\section*{Evaluation of the migrant communities spatial distribution accuracy}

Validation data was extracted from the Continuous Register Statistics of the Municipal Register, regarding the cities of Madrid and Barcelona. The smallest spatial units for this dataset are census tracks, of which the latest available geometrical boundaries for both study areas are the corresponding to 2013. It is well known that census tracks cover all the territory (not only populated areas) and that their size depends on the population density of an area, i.e. the more population density, the smallest the size and vice versa, in order to ensure that all census tracks have a similar number of inhabitants. This means that low density census tracks are larger than those corresponding to the city center, thus integrating non populated territory. For this reason, complementary data about the exact location of the residential areas is needed in order to properly geo-reference population data from census track statistics. In this research, information was extracted from the "Downloads of data and cartography by town" service of the SEC, the point of access to electronic services provided by the Directorate General of Land Registry of Spain. This data was transformed in order to obtain the surface devoted to each land use in each urban parcel. Some data treatment was required in order to obtain the number of people residing in each 500 x 500 m$^2$ grid cell according to the main language spoken in the country of origin. S\ref{TabS9} Table shows the correspondence between country of origin and the languages detected in the main part of this paper. It is important to notice that not all the countries in the world are present in the original table. 

\begin{table}[!h]
\centering
\begin{tabular}{l r}
Language & Country of Origin\\ \hline
German&Germany \\
South Slavic&Bulgaria \\
French&France\\
Italian&Italy\\
West Slavic&Poland\\
Portuguese&Portugal, Brazil\\
English&United Kingdom\\
Romanian&Romania\\
East Slavic&Russia, Ukraine\\
Arabic&Morocco, Algeria\\
Spanish&Spain, Argentina, Bolivia\\
&Colombia, Cuba, Chile, Ecuador\\
&Paraguay, Peru, Dominican Republic\\
&Uruguay, Venezuela\\
Chinese&China\\
Urdu&Pakistan\\
\end{tabular}
\caption{\textbf{Correspondence between languages detected in Twitter users and country of origin.}}
\label{TabS9}
\end{table}

The second step in data treatment was to locate where people in each census track actually live according with the location of residential land. We selected the blocks containing some surface devoted to residential use from the cadastral dataset. With the use of a Geographic Information System (GIS) we were able to intersect these polygons with the census track boundaries, and to assign the population of each census track to its residential land, proportional to the size of each residential polygon within each census track. Finally, the resulting dataset was intersected with the grid used in the previous parts of this research in order to obtain the estimated number of residents of each language in each grid cell. 

Anselin Local Moran’s \textit{I} is a well-known statistic that provides information on the location and size of four types of clusters: a) high-high clusters of significant high values of a variable that are surrounded by high variables of the same variable; b) high-low clusters of significant high values of a variable surrounded by low values of the same variable; c) low-high clusters of significant low values of a variable surrounded by high values of the same variable; and d) low-low clusters of significant low values of a variable surrounded by low values of the same variable. While the typical tools available in most GIS software solutions allow for univariate analysis, GeoDa is an open source product that also allows the computation of bivariate analysis \cite{Anselin2006}, thus enabling the identification of spatial clusters in which high values of one variable are surrounded by high values of the second (i.e. lagged) variable (high-high clusters) and so on.

\begin{table}[h]
\centering
\begin{tabular}{llcccc}
Language & City & \textit{I} & Z-value & pseudo p-value & Spatial Autocorrelation\\ \hline
\multirow{2}{*}{Total} & Barcelona & 0,63 & 236,51 & 0,01 & Positive\\
					   & Madrid & 0,62 & 268,62 & 0,01 & Positive\\
\multirow{2}{*}{Spanish} & Barcelona & 0,62 & 216,99 & 0,01 & Positive\\
						 & Madrid & 0,62 & 267,29 & 0,01 & Positive\\
\multirow{2}{*}{English} & Barcelona & 0,50 & 230,53 & 0,01 & Positive\\
						 & Madrid & 0,38 & 190,62 & 0,01 & Positive\\
\multirow{2}{*}{French} & Barcelona & 0,37 & 151,51 & 0,01 & Positive\\
						& Madrid & 0,32 & 159,25 & 0,01 & Positive\\
\multirow{2}{*}{Italian} & Barcelona & 0,28 & 125,84 & 0,01 & Positive\\
						 & Madrid & 0,26 & 146,32 & 0,01 & Positive\\
\multirow{2}{*}{Portuguese} & Barcelona & 0,32 & 151,20 & 0,01 & Positive\\
							& Madrid & 0,44 & 204,95 & 0,01 & Positive\\
\multirow{2}{*}{Arabic} & Barcelona & 0,08 & 89,88 & 0,01 & Random\\
						& Madrid & 0,07 & 41,50 & 0,01 & Random\\
\multirow{2}{*}{East-Slavic} & Barcelona & 0,21 & 112,83 & 0,01 & Positive\\
							 & Madrid & 0,06 & 37,66 & 0,01 & Random\\
\end{tabular}
\caption{\textbf{Data Validation}. Global Moran's \textit{I}.}
\label{TabS10}
\end{table}

Bivariate global Moran's I (S\ref{TabS10} Table) indicates the existence of positive spatial autocorrelation between the location of tweets and residential areas. In general terms, there is a high positive spatial correlation in both study areas (Moran’s \textit{I} = 0.6). The z and p values have been evaluated through 99 permutations. This value remains high for local language (Spanish in Madrid and Spanish and Catalan in Barcelona). The spatial autocorrelation of foreign languages is a bit lower, which might be in part due to the inconsistencies between Twitter language and available countries of origin in the official statistics (i.e. United Kingdom is the only country of origin for English speakers and so are Morocco and Algeria for Arabic languages). Anyway, Arabic is the only language whose tweets show a random spatial pattern in relation with the location of resident population from Morocco or Algeria in both cities, whereas tweets in English in Barcelona and tweets in Portuguese in Madrid are highly positively spatially correlated with resident population from the UK and Portugal or Brazil, respectively.

\begin{figure}[!h]
\centering
\includegraphics[width=1\linewidth]{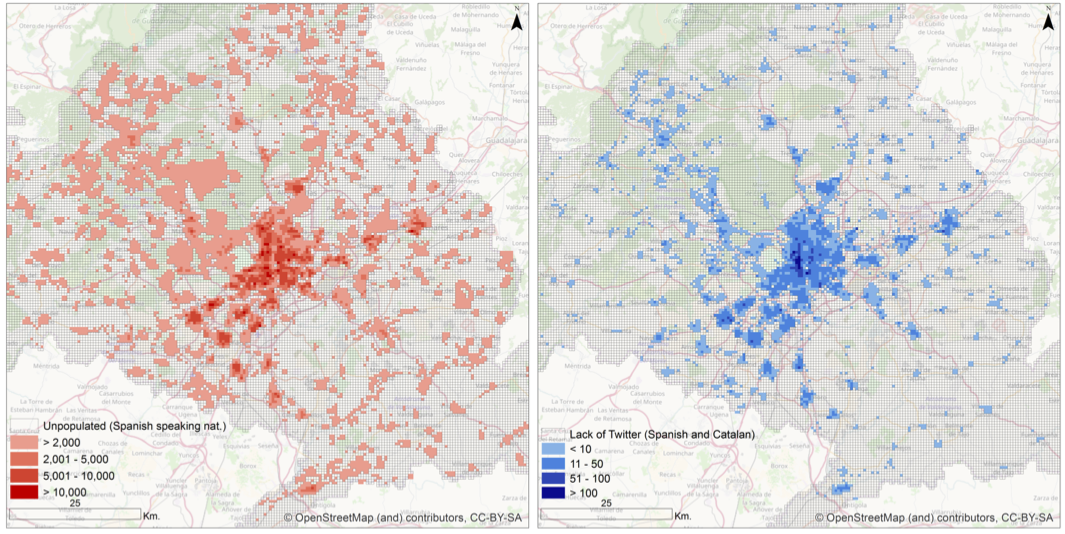}
\caption{\textbf{Data Validation (1/2)}. Distribution of Spanish native users in Madrid, according to official statistics and to our framework of language detection process.}
\label{FigS4}
\end{figure}

\begin{figure}[!h]
\centering
\includegraphics[width=1\linewidth]{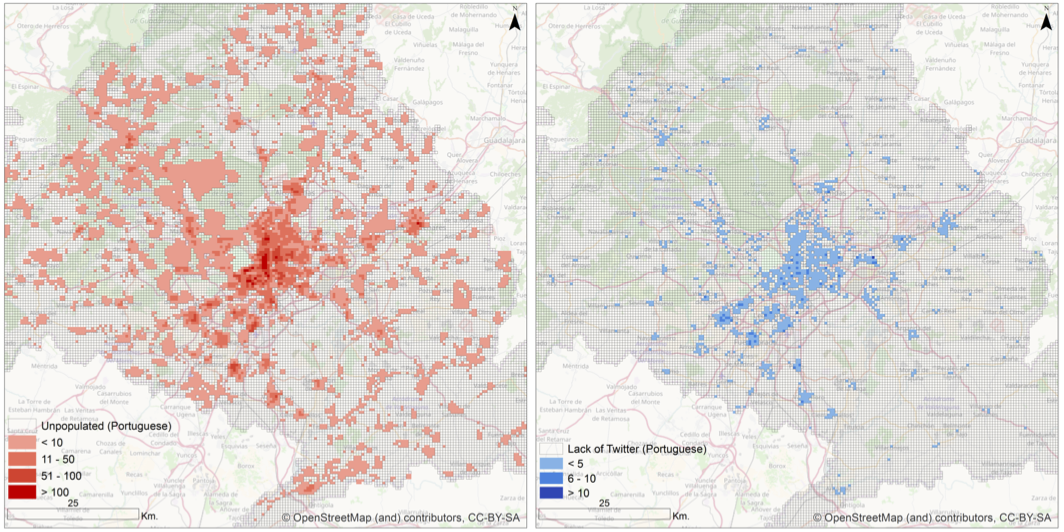}
\caption{\textbf{Data Validation (2/2)}. Distribution of Portuguese native users in Madrid, according to official statistics and to our framework of language detection process.}
\label{FigS5}
\end{figure}

\begin{figure}[!h]
\centering
\includegraphics[width=8cm]{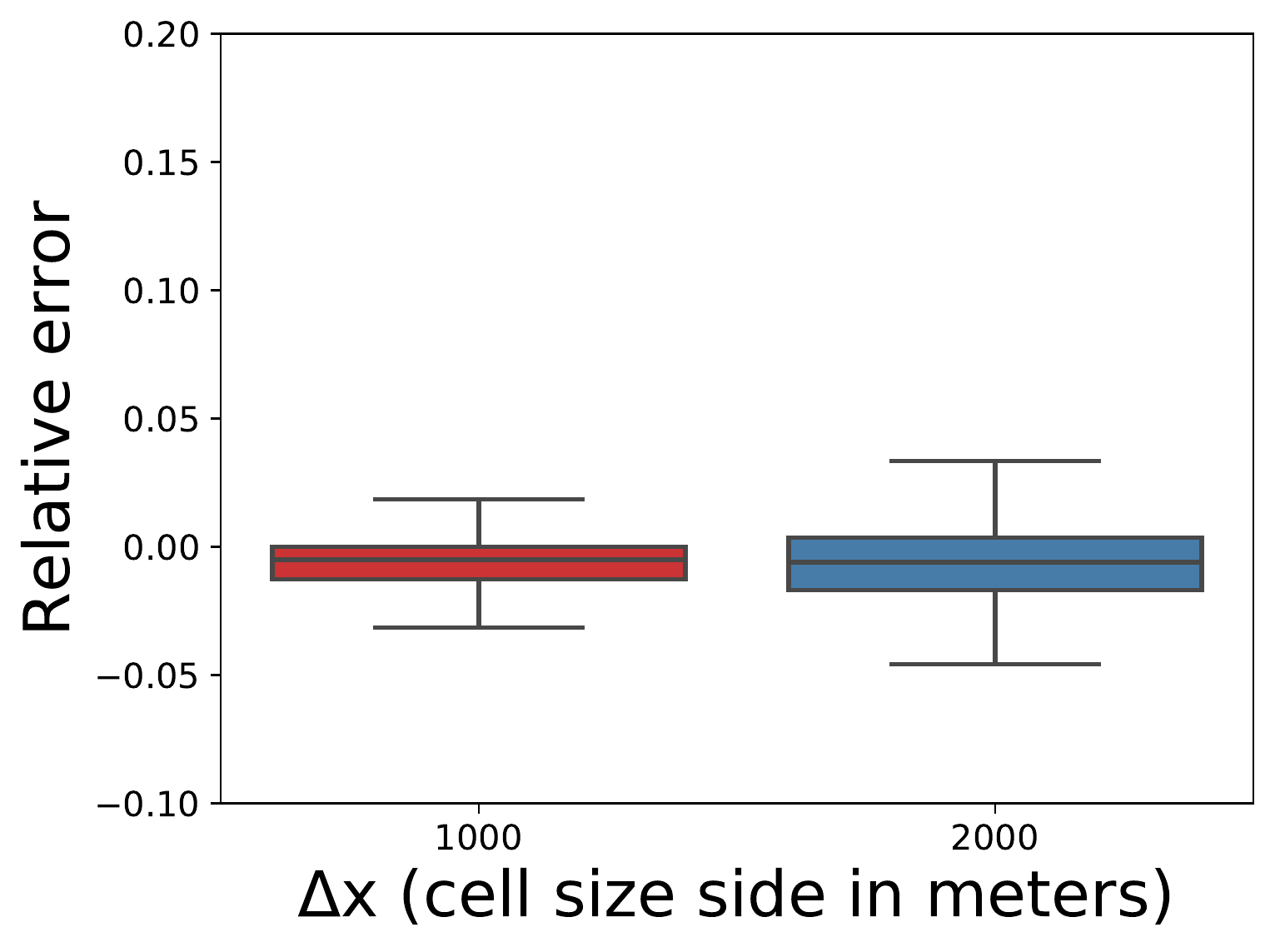}
\caption{\textbf{Relative error of the spatial entropy in function of $\Delta$x}. Box plots of the relative change $\epsilon_{l,c}$ of the link weights in the bipartite spatial integration network taking as reference the unit-like $\Delta$x as the cell side frame of 500 meters, with respect of 4 and 16 times the $\Delta$x for cell side sizes of 1000 and 2000 meters, respectively.}
\label{FigS6}
\end{figure}

\begin{figure}[!h]
\centering
\includegraphics[width=8cm]{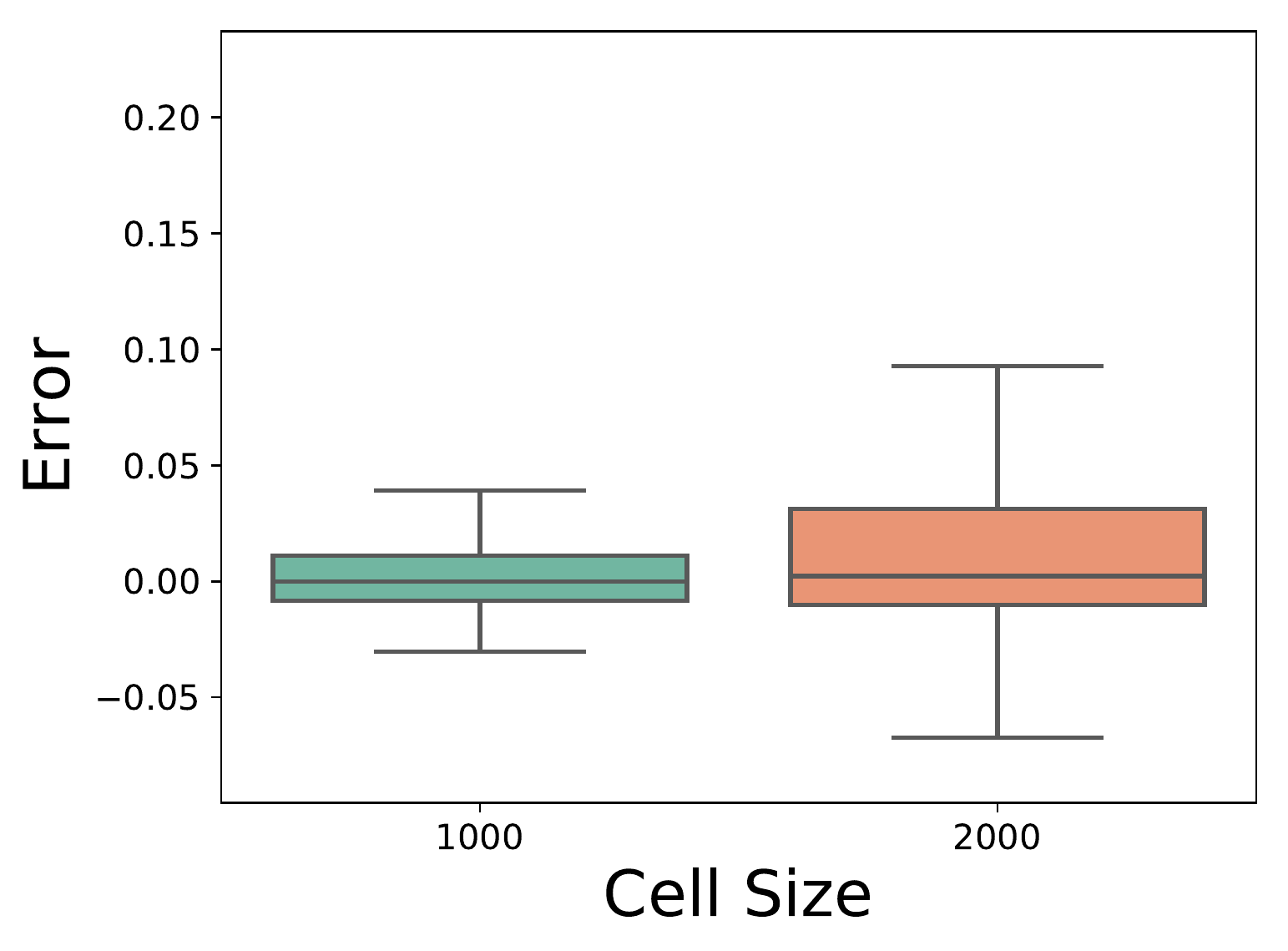}
\caption{\textbf{Relative error of link weights in function of the cell side size}. Box plots of the relative change $\epsilon_{l,c}$ of the link weights in the bipartite spatial integration network taking as reference the 500 meters cell side frame.}
\label{FigS7}
\end{figure}

\clearpage

\end{document}